\documentclass{article} 
\usepackage{amsfonts}
\usepackage{amssymb}

\usepackage{epsfig}
\usepackage{graphicx}

\def\Z{\mathbf{Z}} 
\def\R{\mathbf{R}} 
\def\N{\mathbf{N}}

\def\a{\alpha}
\def\b{\beta}
\def\e{\varepsilon}

\def\G{\Gamma}
\def\Gt{\tilde\G}
\def\l{\lambda}
\def\s{\sigma}

\def\A{{\mathcal A}}

\def\FF{{\mathcal F}} 
 
\def\M{{\mathcal M}} 
\def\L{{\mathcal L}}

\def\V{{\mathcal V}}

\def\pb#1{\left\{#1\right\}}    
\def\PB{\left\{\cdot\,,\cdot\right\}}
\def\inn#1#2{\left\langle#1\,\vert\,#2\right\rangle}

\def\Id{\mathop{\hbox{\rm Id}}\nolimits}

\def\Trace{\mathop{\rm Trace}\nolimits}

\newtheorem{theorem}{Theorem}[section]
\newtheorem{proposition}[theorem]{Proposition}
\newtheorem{remark}[theorem]{Remark}
\newtheorem{lemma}[theorem]{Lemma}

\def \qed{{\hfill $\square$}}

\newenvironment{proof}
  {{\sl Proof}}
  {\par\smallskip}

\newenvironment{equation*}
  {$$}
  {$$}

\newenvironment{eqn*}[1][1.5]
  {$$\renewcommand{\arraystretch}{#1}
      \begin{array}{rcl}}
      {\end{array}$$}

\newenvironment{eqn}[2][1.5]
  {\begin{equation}\label{#2}
   \renewcommand{\arraystretch}{#1}
   \begin{array}{rcl}}
  {\end{array}\end{equation}}

\renewenvironment{matrix}[1]{\left(\begin{array}{#1}}{\end{array}\right)}
\newenvironment{detmat}[1]{\left|\begin{array}{#1}}{\end{array}\right|}

\catcode`\@=11
\def\@nodimen#1{\expandafter\@@nodimen\the#1}%
{\catcode`\p=12\catcode`\t=12\gdef\@@nodimen#1pt{#1}}%
\def\crossed#1{{\setbox0=\hbox{$#1$}%
	\dimen@=\wd0\dimen@i=\ht0\dimen@ii=\dp0
	\advance\dimen@ by 0pt
	\advance\dimen@i by 2pt
	\advance\dimen@ii by 2pt
	\hbox{\special{ps:
}$#1$}}}%
\catcode`\@=12

\def\Liesl#1{\mathfrak{sl}(#1)}     
\def\ds{\displaystyle}
\def\comment#1{}  

\def\mi{{-1}} 
 
\def\p{\partial}
\def\pp#1#2{\frac{\p #1}{\p #2}}
\def\transp#1{{#1^\top}} 
\def\defi#1{\index{#1}\emph{#1}} 

\def\set#1{\left\{#1\right\}}
\def\X{\mathcal{X}} 

\def\bddots{\mathinner{\mkern1mu\raise1pt\vbox{\kern7pt\hbox{.}}\mkern2mu
             \raise4pt\hbox{.}\mkern2mu\raise7pt\hbox{.}\mkern1mu}}

\def\?{(?)\marginpar{|?}}

\begin{document}
\nocite{*}
\centerline{\LARGE Singularity confinement for a class of}
\centerline{\LARGE $m$-th order difference equations of combinatorics}
\bigskip
\centerline{Mark Adler\footnote{Department  of Mathematics, Brandeis University, Waltham, Mass 02454, USA,
adler@brandeis.edu. The support of a National Science Foundation grant \# DMS-04-06287 is gratefully acknowledged},
Pierre van Moerbeke\footnote{D\'epartement de Math\'ematiques, Universit\'e Catholique de Louvain, 1348
Louvain-la-Neuve, Belgium and Brandeis University, Waltham, Mass 02454, USA, vanmoerbeke@math.ucl.ac.be. The
support of a National Science Foundation grant \# DMS-04-06287, a European Science Foundation grant (MISGAM), a
Marie Curie Grant (ENIGMA), Nato, FNRS and Francqui Foundation grants is gratefully acknowledged.} and Pol
Vanhaecke\footnote{The support of a European Science Foundation grant (MISGAM) and a Marie Curie Grant (ENIGMA) is
gratefully acknowledged.}}
%
%
%
\begin{abstract}
In a recent publication, it was shown that a large class of integrals over the unitary group $U(n)$ satisfy
difference equations over $n$, involving a finite number of steps; special cases are generating functions appearing
in questions of longest increasing subsequences in random permutations and words. The main result of the paper
states that these difference equations have the \emph{discrete Painlev\'e property}; roughly speaking, this means
that, after a finite number of steps, the solution to these difference equations may develop a pole (Laurent
solution), depending on the maximal number of free parameters, and immediately after be finite again
(``\emph{singularity confinement}'').  The technique used in the proof is based on an intimate relationship between
the difference equations (discrete time) and the Toeplitz lattice (continuous time differential equations); the
point is that the ``Painlev\'e property'' for the discrete relations is inherited from the ``Painlev\'e property'' of
the (continuous) Toeplitz lattice.
\end{abstract}
\tableofcontents
\
\section{Introduction}
In a recent publication (\cite{AvM2}), we have shown that a large class of integrals over the unitary group ${\bf
U}(n)$ satisfy difference equations over $n$, involving a finite number of steps; these ${\bf U}(n)$-integrals are
motivated by generating functions appearing in questions of longest increasing subsequences in random permutations
and words (see \cite{AvM1}, \cite{AvM2}, \cite{AvM3}, \cite{BR}, \cite{boro}, \cite{rains}, \cite{TW1} and
\cite{TW2}). The main result of the paper, announced in \cite{AvM2}, states that those difference equations, which
are also recursion relations, have the {\em discrete Painlev\'e property}; roughly speaking, this means that the
solution to these difference equations may develop a pole (formal Laurent solution) after a finite number of steps
and immediately after be finite again. Moreover, these formal Laurent solutions depend on the \emph{maximal}
number of free parameters, which equals ((order of difference equation) $-1$) $ \times$ ($\dim$ of phase~space),
with the poles disappearing after a finite number of steps (``\emph{singularity confinement}'').

The technique used in the proof is new and is based on an intimate relation between the difference equations
(discrete time) and the Toeplitz lattice (continuous time differential equations), introduced in \cite{AvM1}; the
point is that the the ``Painlev\'e property'' for the discrete relations are inherited from the ``Painlev\'e
property'' of the (continuous) Toeplitz lattice. Before making a more precise statement and describing the
technique, recall the basic facts about the Toeplitz lattice and the recursion relations \cite{AvM1,AvM2}.


\smallskip

For $k\in\N $ and $\epsilon\in\{-1,0,1\}$, consider the matrix integrals
\begin{equation}\label{matrix integral}
  \tau_k^\epsilon(t,s)=\int_{{\bf U}(k)}
  (\det M)^{\epsilon+\gamma} e^{\sum_{j=1}^\infty\Trace(t_jM^j-s_j M^{-j})}\,dM
\end{equation}%
where $dM$ is Haar measure on ${\bf U}(k)$. Special choices of $t_j$
and $s_j$ lead to generating functions in combinatorics (see \cite{AvM2}). Set $\tau:=\tau^0$ and
$\tau^\pm:=\tau^{\pm1}$. In \cite{AvM2} it was shown that the ratios
\begin{equation*}
  x_k(t,s):=(-1)^k\frac{\tau_k^+(t,s)}{\tau_k(t,s)},
  \qquad y_k(t,s):=(-1)^k\frac{\tau_k^-(t,s)}{\tau_k(t,s)},
\end{equation*}%
satisfy the {\em Toeplitz lattice}, an integrable Hamiltonian system,
\begin{equation}\label{toeplitz_intro}
\renewcommand{\arraystretch}{3}
\begin{array}{rclrcl}
  \ds\pp{x_k}{t_i}&=&\ds(1-x_ky_k)\pp{H_i^{(1)}}{y_k},\qquad &\ds\pp{y_k}{t_i}&=&-\ds(1-x_ky_k)\pp{H_i^{(1)}}{x_k},\\
  \ds\pp{x_k}{s_i}&=&\ds(1-x_ky_k)\pp{H_i^{(2)}}{y_k},\qquad &\ds\pp{y_k}{s_i}&=&-\ds(1-x_ky_k)\pp{H_i^{(2)}}{x_k}.
\end{array}
\end{equation}
and moreover, $\tau_n$ is a polynomial expression in the variables $x_k$ and $y_k$ and $\tau_1$:
$$\tau_n=\tau_1^n\prod_{k=1}^{n-1}(1-x_ky_k)^{n-k}.$$
The Hamiltonians $H_i^{(l)}$ appearing in (\ref{toeplitz_intro}) are given by
\begin{equation*}
   H_i^{(l)}=-\frac1i\Trace L_l^i,\qquad i=1,2,3,\dots,\ l=1,2,
\end{equation*}%
where the  matrices $L_1$ and $L_2$ are defined by
  \begin{equation}
   L_1:=
   \left(\begin{tabular}{lllll}
                  $-x_1y_0$  &  $1-x_1y_1$ & ~~ $0$      & ~~ $0$ &   \\
                  $-x_2y_0$ &  $-x_2y_1$ & $1-x_2y_2$& ~~ $0$   & \\
                  $-x_3y_0$ &  $-x_3y_1$ & $ -x_3y_2$&  $1-x_3y_3$ & \\
                  $ -x_4y_0$ &  $ -x_4y_1$ & $-x_4y_2$  & $ -x_4y_3$ &\\
                  & &  &    &  $\ddots$\\
           \end{tabular}\right)
 \end{equation}
and
  \begin{equation}
   L_2:= \left(\begin{tabular}{lllll}
                         $-x_0y_1$  &  $-x_0y_2$ & $-x_0y_3$     & $-x_0y_4$ &\\
                         $1 -x_1y_1$ &  $-x_1y_2$  & $-x_1y_3$& $-x_1y_4$& \\
                         ~~$0$       &  $1 -x_2y_2$ & $ -x_2y_3$&$-x_2y_4$ & \\
                         ~~$0$       &  ~~$0$      & $ 1
                    -x_3y_3$  & $ -x_3y_4$   &  \\
                         & &  &    &  $\ddots$\\
                  \end{tabular}\right).
  \end{equation}
The system admits a reduction, interesting in its own right, obtained by putting $x_k=y_k$ for all $k$. We refer to
it as the \emph{self-dual Toeplitz lattice}.

\smallskip

In \cite{AvM1}, it was shown that the matrix integrals (\ref{matrix integral}) satisfy a $\Liesl{2,\R}$-algebra of
Virasoro constraints, which combined with the Toeplitz lattice equations, lead to difference equations for $x_k$
and $y_k$ given in \cite{AvM2}, a subset of the cases leading to recursion relations, which we now describe. Given
arbitrary polynomials%
\begin{equation*}
  P_1(\lambda):=\sum_{i=1}^{N }\frac{u_i\lambda^i}i,
  \qquad\hbox{and}\qquad P_2(\lambda)
  :=\sum_{i=1}^{N }\frac{u_{-i}\lambda^i}i,
\end{equation*}%
the variables
\begin{equation*}
  x_k(t,s):=(-1)^k\frac{\tau_k^+(t,s)}{\tau_k(t,s)},
  \qquad y_k(t,s):=(-1)^k\frac{\tau_k^-(t,s)}{\tau_k(t,s)},
\end{equation*}
with 
$$\tau_k^\epsilon(t,s)=\int_{{\bf U}(k)}
  (\det M)^{\epsilon+\gamma}
  e^{\Trace(P_1(M)-P_2(M^\mi))}\,dM,
$$
%
%
satisfy $2N+1$ step difference equations in terms of the
matrices $L_1$ and $L_2$ defined above, (set
$v_i:=1-x_iy_i$)
%
\begin{equation}
  \renewcommand{\arraystretch}{3}
  \begin{array}{rcl}
    \Gamma_k(x,y)&:=&
 \displaystyle\frac{v_k}{y_k}\left(
    \renewcommand{\arraystretch}{1.3}
    \begin{array}{c}
      -(L_1P'_1(L_1))_{k+1,k+1}-(L_2P_2'(L_2))_{k,k}\\
      +(P'_1(L_1))_{k+1,k}+(P_2'(L_2))_{k,k+1}
    \end{array}
    \right)+kx_k=0,\\
     \tilde\Gamma_k(x,y)&:=&
 \displaystyle\frac{v_k}{x_k}\left(
    \renewcommand{\arraystretch}{1.3}
    \begin{array}{c}
      -(L_1P'_1(L_1))_{k,k}-(L_2P_2'(L_2))_{k+1,k+1}\\
      +(P'_1(L_1))_{k+1,k}+(P_2'(L_2))_{k,k+1}
    \end{array}
    \right)+ky_k=0.
 \label{a} \end{array}
\end{equation}
Looking closely, one observes that these difference equations $\Gamma_k$ and~$\tilde\Gamma_k$ are indeed linear in
$x_{k+N}$ and $y_{k+N}$, and can thus be solved in terms of $x_{k-N},\,y_{k-N},$ $\dots,x_{k+N-1},\,y_{k+N-1}$.
%
%
See the appendix for a proof of this fact.

This paper deals with the difference equations (\ref{a}) for their own sake, without further reference to the
special solution $x_k(t,s)$ and $y_k(t,s)$, given by the unitary matrix integrals above. Moreover, we will consider
the bi-infinite Toeplitz lattice, which is defined as in~(\ref{toeplitz_intro}), but with $k\in\Z$. The recursion
relations are then also considered for $k\in\Z$, with the semi-infinite case obtained by specialization. The
bi-infinite Toeplitz lattice will be introduced in Section \ref{toeplitz_sec}, where we also discuss the self-dual
Toeplitz lattice and the recursion relations.

It came as a surprise that the generic solutions of these (very general) equations (\ref{a}) have the {\em
singularity confinement} property (see \cite{GNR} and \cite{Su}); a fact, which had been observed by Borodin (see
\cite{boro}) in the very special case of unitary matrix integrals related to longest increasing sequences of random
permutations. The main result of the paper is to show this surprising fact for the difference equations (\ref{a}),
namely:

\begin{theorem}[singularity confinement: general case]\label{intro_thm}
  For any $n\in\Z$, the difference equations $\Gamma_k(x,y)=\tilde\Gamma_k(x,y)=0,~ (k\in \Z)$ admit a formal
  Laurent solution $x=(x_k(\lambda))_{k\in\Z}$ and $y=(y_k(\lambda))_{k\in\Z}$ in a parameter $\lambda$, having a
  (simple) pole at $k=n$ and $\lambda= 0$, and no other singularities. These solutions depend on $4N$ non-zero free
  parameters
  $$\alpha_{n-2N},
  \dots,\alpha_{n-2},\alpha_{n-1},\beta_{n-2N},
  \dots,\beta_{n-2}~\mbox{and}~\lambda. $$
%
%
  Setting $z_n:=(x_n,y_n)$ and $\gamma_i:=(\alpha_i,\beta_i)$, and $\gamma:=(\gamma_{n-2N}, \dots,\gamma_{n-2},
  ~\alpha_{n-1})$, the explicit series with coefficients rational in $\gamma$ read as follows:
\begin{equation*}
  \renewcommand{\arraystretch}{1.7}
  \begin{array}{rclrcl}
  z_k(\lambda)&=&\sum_{i=0}^\infty z_k^{(i)}
  (\gamma)\lambda^i,\qquad &k&<&n-2N,\\
  z_k(\lambda)&=&\gamma_k,&n-2N&\leq& k\leq n-2,\\
  x_{n-1}(\lambda)&=&\alpha_{n-1},\\
  y_{n-1}(\lambda)&=&1/\alpha_{n-1}+\lambda,\\
  z_n(\lambda)&=&
    \frac1\lambda\sum_{i=0}^\infty z_n^{(i)}(\gamma)\lambda^i,\\
  z_k(\lambda,\gamma)&=&
   \sum_{i=0}^\infty z_k^{(i)}(\gamma)\lambda^i,\qquad &n&<&k.
  \end{array}
\end{equation*}%
\end{theorem}

\noindent For the self-dual case, the statement reads as
follows:

\begin{theorem}[singularity confinement: self-dual case]
 For any $n\in\Z$, the difference equations $\Gamma_k(x)=0,~ (k\in \Z)$ admit two\footnote{parametrized by
  $\epsilon=\pm1.$} formal Laurent solution $x=(x_k(\lambda))_{k\in\Z}$ in a parameter $\lambda$, having a
  (simple) pole at $k=n$ only and $\lambda= 0$. These solutions depend on $2N$ non-zero free parameters
  $$\alpha=(\alpha_{n-2N},\dots,\alpha_{n-2})~\mbox{and}
  ~\lambda$$
%
%
Explicitly, these series with coefficients rational in $\alpha$ are given by
\begin{equation*}
  \renewcommand{\arraystretch}{1.7}
  \begin{array}{rclrcl}
  x_k(\lambda)&=&\sum_{i=0}^\infty x_k^{(i)}(\alpha)\lambda^i,\qquad &k&<&n-2N,\\
  x_k(\lambda)&=&\alpha_k,&n-2N&\leq& k\leq n-2,\\
  x_{n-1}(\lambda)&=&\e+\lambda,\\
  x_n(\lambda)&=&\frac1\lambda\sum_{i=0}^\infty
  x_n^{(i)}(\alpha)\lambda^i,\\
  x_{n+1}(\lambda)&=&-\e+\sum_{i=1}^\infty x_{n+1}^{(i)}
  (\alpha)\lambda^i,\qquad\\
  x_k(\lambda)&=&\sum_{i=0}^\infty x_k^{(i)}(\alpha)
  \lambda^i,\qquad &n+1&<&k.
  \end{array}
\end{equation*}%
\end{theorem}

\smallskip

The proof of Theorems 1.1 and 1.2 is by no means direct, but proceeds via the Painlev\'e analysis for the Toeplitz
lattice. As a starting point,
the zero locus ${\cal M},$ of all polynomials $\Gamma_k$ and $\tilde\Gamma_k,$ form an invariant manifold for the
vector field of the Toeplitz lattice with Hamiltonian $H_1^{(1)}-H_2^{(2)}$, by viewing the coefficients of
$P_1(\lambda)$ and $P_2(\lambda)$ as constants, except for $u_{\pm 1}$, which moves linearly in time.  Explicitly,
this vector field is given by
\begin{equation}\label{toeplitz_vf_intro}
\renewcommand{\arraystretch}{2}
  \begin{array}{rcl}
    \displaystyle \pp{x_k}t&=&(1-x_ky_k)(x_{k+1}-x_{k-1}),\\
    \displaystyle \pp{y_k}t&=&(1-x_ky_k)(y_{k+1}-y_{k-1}),
  \end{array}
  \qquad \qquad k\in\Z.
\end{equation}
In the self-dual case, this vector field reduces to
\begin{equation}\label{sd_toeplitz_vf_intro}
  \pp{x_k}t=(1-x^2_k)(x_{k+1}-x_{k-1}),\qquad \qquad k\in
  \Z.
\end{equation}
The first idea is then to restrict the principal balances (formal Laurent solutions depending on the maximal number
($=\dim$ phase space $-1$) of free parameters, besides time) of (\ref{toeplitz_vf_intro}) to these invariant
manifolds. We fix $n$ and look for a formal Laurent solution to the Toeplitz lattice that has a (simple) pole for
$x_n$ and $y_n$ only, and we find a unique such family, as given by the following proposition:
\begin{proposition}\label{lau_prop_intro}
For arbitrary but fixed $n$, the first Toeplitz lattice vector field (\ref{toeplitz_vf_intro}) admits the following
formal Laurent solutions,
\begin{eqnarray*}
  x_n(t)&=&\frac{1}{(a_{n-1}-a_{n+1})t}
  \Bigl(a_{n-1}a_{n+1}(1+at)+O(t^2)\Bigr)  \\
  y_n(t)&=&\frac{1}{(a_{n-1}-a_{n+1})t}
 \left( -1+\bigl(a+\frac{a_{n+1}a_+-a_{n-1}a_-}
 {a_{n+1}-a_{n-1}}\bigr)t+O(t^2)\right)
  \\  \\
  x_{n\pm 1}(t)&=&a_{n\pm1}+a_{n\pm1}a_\pm t+O(t^2)\\
  y_{n\pm 1}(t)&=& 1/a_{n\pm1}-a_\pm/a_{n\mp1}t+O(t^2)
\end{eqnarray*}
whereas for all remaining $k$ such that $|k-n|\geq 2$,
\begin{eqnarray}\label{lau_sol_intro}
 x_k(t)&=&a_k+(1-a_kb_k)(a_{k+1}-a_{k-1})t+O(t^2)\\
  y_k(t)&=&b_k+(1-a_kb_k)(b_{k+1}-b_{k-1})t+O(t^2)
\end{eqnarray}
  where $a,\,a_\pm,\,a_{n\pm1}$ and all
  $a_i,~b_i$, with
  $i\in\Z\setminus\{n-1,n,n+1\}$ and with
  $b_{n\pm 1}=1/a_{n\pm 1}$, are arbitrary free parameters, and with
  $(a_{n-1}-a_{n+1})a_{n-1}a_{n+1}\neq0$. In the self-dual case it admits the
  following two formal Laurent solutions, parametrized by
  $\e=\pm1$,
  \begin{eqnarray}\label{sd_lau_sol_intro}
   x_n(t)&=&-\frac{\e}{2t}\left(1+(a_+-a_-)t+O(t^2)\right),\nonumber\\
    x_{n\pm1}(t)&=&\e\left(\mp1+4a_\pm t+O(t^2)\right),\\
    x_k(t)&=&\e\left(a_k+(1-a_k^2)(a_{k+1}-a_{k-1})t
    +O(t^2)\right),\qquad\quad \vert k-n\vert\geq 2,\nonumber\\
  \end{eqnarray}
  where $a_+,a_-$ and all $a_i$, with $i\in\Z\setminus\set{n-1,n,n+1}$ are arbitrary free parameters and
  $a_{n-1}=-a_{n+1}=1$.
\end{proposition}
Together with time $t$ these parameters are in bijection with the phase space variables; we can put for the general
Toeplitz lattice for example $z_k\leftrightarrow (a_k,b_k)$ for $\vert k-n\vert\geq1$ and $x_{n\pm1}\leftrightarrow
a_{n\pm1}$ and $y_{n\pm1},x_n,y_n\leftrightarrow a_\pm,a,t$. Thus, this formal Laurent solution is the natural
candidate to work with; see Section~\ref{laurent_sec}.

It is however, a priori, not clear that these formal Laurent solutions can be restricted to the invariant manifold
$\cal M$. Indeed, upon introducing a proper time-dependence for $u$ already mentioned, one has that
$\Gamma_k(t):=\Gamma_k(x(t),y(t);u(t))$ and $\tilde\Gamma_k(t):=\tilde\Gamma_k(x(t),y(t);u(t))$ satisfy a system of
differential equations, as given in the following proposition:
\begin{proposition}\label{diff_eq_intro}
  Upon setting $\frac{\p u_{\pm i}}{\p t}=\delta_{1i},$ the recursion relations satisfy the following differential equations
  \begin{equation}\label{G_diff_eq_intro}
  \renewcommand{\arraystretch}{1.3}
    \begin{array}{rcl}
      \dot\Gamma_k&=&v_k(\Gamma_{k+1}-\Gamma_{k-1})+(x_{k+1}-x_{k-1})(x_k\Gt_k-y_k\G_k),\\
      \dot{\tilde\Gamma}_k&=&v_k(\tilde\Gamma_{k+1}-\tilde\Gamma_{k-1})-(y_{k+1}-y_{k-1})(x_k\Gt_k-y_k\G_k),
    \end{array}
  \end{equation}
  which specialize in the self-dual case (\ref{sd_toeplitz_vf_intro}) to
  \begin{equation}\label{G_sd_diff_eq_intro}
  \renewcommand{\arraystretch}{1.3}
    \begin{array}{rcl}
      \dot\Gamma_k&=&v_k(\Gamma_{k+1}-\Gamma_{k-1}).\\
    \end{array}
  \end{equation}
\end{proposition}

In addition to Propositions 1.3 and 1.4, many other arguments are needed to fine-tune the free parameters, when
going from the Laurent solutions of the Toeplitz lattice to the existence of formal Laurent solutions to the
difference equations, depending on the announced number of free parameters. See Section \ref{conf_general_sec}. The
proof of these facts will be spread over two sections, as the arguments get rather involved; see Section
\ref{conf_self-dual_sec} for the self-dual case and Section \ref{conf_general_sec} for the case of the general
Toeplitz lattice.
This ultimately leads to the proof of the main Theorems 1.1 and 1.2.

%
\section{An invariant manifold $\M$ for the first Toeplitz flow}
\label{toeplitz_sec}
In this section we introduce the bi-infinite Toeplitz lattice, in analogy with the semi-infinite Toeplitz lattice,
introduced in \cite{AvM1}. We also recall the basic formulas related to the invariant manifold $\M$ that we will
introduce below (see \cite{AvM2}).

The (bi-infinite) \defi{Toeplitz lattice} consists of two infinite strings of vector fields on the (real or complex)
linear space of bi-infinite sequences $(x_i,y_i)_{i\in\Z}$. The particular vector field that we will be interested in
(the ``first'' Toeplitz vector field) is given by
\begin{equation}\label{toeplitz_vf}
\renewcommand{\arraystretch}{2}    
  \begin{array}{rcl}
    \displaystyle \pp{x_k}t&=&(1-x_ky_k)(x_{k+1}-x_{k-1}),\\
    \displaystyle \pp{y_k}t&=&(1-x_ky_k)(y_{k+1}-y_{k-1}),
  \end{array}
  \qquad \qquad k\in\Z.
\end{equation}
The semi-infinite Toeplitz lattice is obtained from it by setting $(x_k,y_k)=(0,0)$ for $k<0$ and $(x_0,y_0)=(1,1)$. The
invariant polynomials of the matrices $L_1$ and $L_2$, defined by
\begin{equation}\label{lax_mats}
  \renewcommand{\arraystretch}{1.3} 
  \begin{array}{rcl}
    (L_1)_{ij}&:=&\left\{
      \begin{array}{ccl}
        -x_iy_{j-1}+\delta_{i+1,j}\ &\hbox{ if } &j-i\leq1,\\
        0&\hbox{ if } &j-i>1,\\
      \end{array}
     \right.\\
    (L_2)_{ij}&:=&\left\{
      \begin{array}{ccl}
        -y_jx_{i-1}+\delta_{j+1,i}&\hbox{ if } &j-i\geq1,\\
        0&\hbox{ if } &j-i<1,\\
      \end{array}
     \right.
  \end{array}
\end{equation}
provide two infinite strings of constants of motion $H_i^{(1)}$ and $H_i^{(2)}\ (i\in\Z)$ of (\ref{toeplitz_vf}),
defined by
\begin{equation}
  H_i^{(l)}=-\frac1i\Trace L^i_l,\qquad i=1,2,3,\dots,\quad l=1,2.
\end{equation}
The first Toeplitz vector field (\ref{toeplitz_vf}) is the Hamiltonian vector field that corresponds to
\begin{equation*} 
  H_1:=H_1^{(1)}-H_1^{(2)}=\Trace (L_2-L_1)=\sum_{i\in\Z}(x_iy_{i-1}-x_{i-1}y_i),
\end{equation*}%
with respect to the Poisson structure defined by
\begin{equation*}
  \pb{x_i,x_j}=\pb{y_i,y_j}=0,\qquad \pb{x_i,y_j}=(1-x_iy_j)\delta_{ij},
\end{equation*}%
and the functions $H_i^{(1)}$ and $H_i^{(2)}$ are all in involution with respect to $\PB$, as follows from a direct
computation. As a corollary, all Hamiltonian vector fields $\X_i^{(1)}:=\pb{\cdot\,,H_i^{(1)}}$ and
$\X_i^{(2)}:=\pb{\cdot\,,H_i^{(2)}}$ commute. If we denote $\inn AB:=\Trace AB$, whenever this makes sense, then for
$i=1,2,\dots,$
\begin{equation*}
  \X_i^{(1)}[x_k]=\pb{x_k,-\frac1i\Trace L_1^i}=-(1-x_ky_k)\inn{L_1^{i-1}}{\pp{L_1}{y_k}},
\end{equation*}%
and similarly for $\X_i^{(1)}[y_k]$, which leads to the following expression for the vector field $\X_i^{(1)}$,
\begin{equation}\label{one_vec}
  \X_i^{(1)}:\left\{
\renewcommand{\arraystretch}{2}
   \begin{array}{rcl}
      \displaystyle \pp{x_k}{t_i}&=&-(1-x_ky_k)\inn{L_1^{i-1}}{\pp{L_1}{y_k}},\\
      \displaystyle \pp{y_k}{t_i}&=&(1-x_ky_k)\inn{L_1^{i-1}}{\pp{L_1}{x_k}}.
   \end{array} 
  \right.
\end{equation}
The vector field $\X_i^{(2)}$, has the same form, but with $L_1$ replaced by $L_2$. This is a particular case of a
phenomenon that we will refer to as duality. Namely, there is a natural automorphism $\s$ of our phase space, given by
$\s:(x_i,y_i)_{i\in\Z}\mapsto (y_i,x_i)_{i\in\Z}$. It preserves the first Toeplitz vector field~(\ref{toeplitz_vf}), it
permutes the Hamiltonians $H_i^{(1)}\leftrightarrow H_i^{(2)}$, it permutes the Lax operators as follows:
$L_1\leftrightarrow\transp{L_2}$ and it reverses the sign of the Poisson structure. The first Toeplitz vector field
(\ref{toeplitz_vf}) can be restricted to the fixed point locus $(x_i=y_i)_{i\in\Z}$ of $\s$, which leads to the
\defi{self-dual} (bi-infinite) Toeplitz lattice,
\begin{equation}
  \pp{x_k}t=(1-x_k^2)(x_{k+1}-x_{k-1}),\qquad k\in\Z.
\end{equation}
All constructions in this paper will be done for this self-dual lattice first, and then for the general Toeplitz
lattice. This is not only for pedagogical reasons: even if the ideas that lead to the proofs are similar in both cases,
the self-dual lattice can for our purposes not be treated as a particular case of the general Toeplitz lattice, as we
will see.

For $i=1$, the equations (\ref{one_vec}) for $\X_i^{(1)}$ and for $\X_i^{(2)}$ specialize to
\begin{equation}\label{one_vec_bis}
\renewcommand{\arraystretch}{1.6}
  \begin{array}{rcl}
    \X_1^{(1,2)}[x_k]&=&(1-x_ky_k)x_{k\pm1},\\
    \X_1^{(1,2)}[y_k]&=&-(1-x_ky_k)y_{k\mp1}.
  \end{array} 
\end{equation}

Fixing $2N$ constants $u:=(u_{-N},\dots,u_{-1},u_1,\dots,u_N)$, with $u_{N}\neq0$ and $u_{-N}\neq0$, we consider the
polynomials
\begin{equation}\label{P_def}
  P_1(\l):=\sum_{i=1}^{N}\frac{u_i\l^i}i, \qquad\hbox{and}\qquad P_2(\l):=\sum_{i=1}^{N}\frac{u_{-i}\l^i}i.
\end{equation}
They lead to two strings of polynomials\footnote{The structure of the matrices $L_1$ and $L_2$ implies that $\G_k$ and
$\Gt_k$ are indeed polynomials. They are also polynomials (of degree 1) in the variables $u_i$, but we often do not
mention this, because we think of these variables as parameters.}  $\G_k$ and $\Gt_k$ in $x_i,\,y_i\ (i\in\Z)$, where
$v_k:=1-x_ky_k$ ($=\s(v_k)$):
\begin{equation}\label{gam_def}
  \renewcommand{\arraystretch}{3}
  \begin{array}{rcl}
    \G_k(x,y;u)&:=&\displaystyle\frac{v_k}{y_k}\left(
    \renewcommand{\arraystretch}{1.3}
    \begin{array}{c}
      -(L_1P'_1(L_1))_{k+1,k+1}-(L_2P_2'(L_2))_{k,k}\\
      +(P'_1(L_1))_{k+1,k}+(P_2'(L_2))_{k,k+1}
    \end{array}
    \right)+kx_k,\\
    \Gt_k(x,y;u)&:=&\displaystyle\frac{v_k}{x_k}\left(
    \renewcommand{\arraystretch}{1.3}
    \begin{array}{c}
      -(L_1P'_1(L_1))_{k,k}-(L_2P_2'(L_2))_{k+1,k+1}\\
      +(P'_1(L_1))_{k+1,k}+(P_2'(L_2))_{k,k+1}
    \end{array}
    \right)+ky_k.
  \end{array}
\end{equation}
Notice that the only elements that appear in these polynomials are the diagonal and next-to-diagonal entries of $L^l_1$
and $L^l_2$ for $l=1,\dots,N$.  For fixed $u$ we consider the zero locus of all polynomials $\G_k$ and $\Gt_k$,
\begin{equation}\label{Mu}
  \M_u:=\bigcap_{k\in \Z}\set{(x_i,y_i)_{i\in\Z}\mid
  \G_k(x,y;u)=0\hbox{ and } \Gt_k(x,y;u)=0}.
\end{equation}
In terms of the variables $x_i$ and $y_i$ the leading terms of $\G_k$ and $\Gt_k$ are given by
\begin{eqnarray*}
  \G_k(x,y;u)&=&u_Nx_{k+N}\prod_{i=0}^{N-1}v_{k+i}+\cdots+u_{-N}x_{k-N}
         \prod_{i=0}^{N-1}v_{k-i},\\
  \Gt_k(x;y;u)&=&u_{-N}y_{k+N}\prod_{i=0}^{N-1}v_{k+i}+\cdots+u_{N}y_{k-N}
         \prod_{i=0}^{N-1}v_{k-i}.
\end{eqnarray*}
See the Appendix for a precise statement, a few more terms and a proof.  We often write $\Delta_k$ as a shorthand for
the vector $\transp{(\G_k,\Gt_k)}$ and $z_k$ for $\transp{(x_k,y_k)}$.

\smallskip

In order to get the corresponding formulas for the self-dual case we put $\sigma(u_i):=u_{-i}$, so that $\s$ permutes
$P_1$ and $P_2$, as well as $\G_k$ and $\Gt_k$, hence $P_1=P_2$ in the self-dual case, and $\G_k=\Gt_k$. Writing
$L:=L_1$ and $P:=P_1$, the polynomials $\G_k$ and $\Gt_k$ reduce in that case to
\begin{equation}\label{gam_def_selfdual}
   \G_k(x;u):=\frac{v_k}{x_k} \Bigl(2(P'(L))_{k+1,k}-(LP'(L))_{k+1,k+1}-(LP'(L))_{k,k}\Bigr)+kx_k,
\end{equation}
while its leading terms are now given by 
\begin{equation}\label{high_selfdual}
  \G_k(x;u)=u_Nx_{k+N}\prod_{i=0}^{N-1}v_{k+i}+\cdots+u_{N}x_{k-N} \prod_{i=0}^{N-1}v_{k-i}.
\end{equation}
The zero locus $\M_u$ now takes the simple form
\begin{equation}\label{Mu_self_dual}
  \M_u:=\bigcap_{k\in \Z}\set{(x_i)_{i\in\Z}\mid \G_k(x;u)=0}.
\end{equation}

Following (\cite{AvM2}) we show that, upon introducing a proper time dependence, the polynomials $\G_k$ and $\Gt_k$
satisfy a simple set of differential equations, showing that the zero locus (\ref{Mu}) of these polynomials is a
(time-dependent) invariant manifold of the first Toeplitz flow (\ref{toeplitz_vf}).
\begin{proposition}
  Let $(x(t),y(t))$ be a solution to the first Toeplitz vector field (\ref{toeplitz_vf}), to wit:
  \begin{equation*}
    \pp{}{t}
      \begin{matrix}{c}
        x(t)\\y(t)
      \end{matrix}=
      \left(\X_1^{(1)}-\X_1^{(2)}\right)
      \begin{matrix}{c}
        x(t)\\y(t)
      \end{matrix}, 
  \end{equation*}%
  and let $\G(t):=\G(x(t),y(t);u(t))$ and $\Gt(t):=\G(x(t),y(t);u(t))$, where
  \begin{equation}\label{u_def}
    u(t)=(u_{-N},\dots,u_{-2},u_{-1}+t,u_1+t,u_2,\dots,u_N).
  \end{equation}
  Then $\G(t)$ and $\Gt(t)$ satisfy the following differential equations:
\begin{equation}\label{G_diff_eq}
\renewcommand{\arraystretch}{1.3}
  \begin{array}{rcl}
    \dot\G_k&=&v_k(\G_{k+1}-\G_{k-1})+(x_{k+1}-x_{k-1})(x_k\Gt_k-y_k\G_k),\\
    \dot{\Gt}_k&=&v_k(\Gt_{k+1}-\Gt_{k-1})-(y_{k+1}-y_{k-1})(x_k\Gt_k-y_k\G_k).
  \end{array}
\end{equation}
  In particular, $\M_{u(t)}$ is a (time-dependent) invariant manifold of the first Toeplitz flow.
  In the self-dual case, these differential equations specialize to
  \begin{equation}\label{G_diff_eq_selfdual}
    \dot\G_k=v_k(\G_{k+1}-\G_{k-1}).
  \end{equation}
  Then $\M_{u(t)}$ is a (time-dependent) invariant manifold of the first vector field of the self-dual Toeplitz lattice,
  where $u(t)=(u_1+t,u_2,\dots,u_N)$.
\end{proposition}
\begin{proof}
We first show that 
\begin{equation}\label{G_eq}
\renewcommand{\arraystretch}{1.3} 
  \begin{array}{rcl}
    \G_k(x,y;u)&=&\V^u[x_k]+kx_k,\\
    \Gt_k(x,y;u)&=&-\V^u[y_k]+ky_k,
  \end{array}
\end{equation}
where $\V^u$ is the Hamiltonian vector field
\begin{equation*}
  \V^u:=\sum_{i=1}^N\left(u_i\X_i^{(1)}+u_{-i}\X_i^{(2)}\right).
\end{equation*}%
It suffices to prove that $\G_k(x,y;u)=\V^u[x_k]+kx_k,$ the other identity being obtained by duality (indeed,
$\s(\V^u)=-\V^u$ since $\s(\X_i^{(1)})=-\X_i^{(2)}$). In view of the Definition (\ref{gam_def}) of $\G_k$ this means
that we need to prove that
\begin{eqn}[2]{to_pr}
  \X_i^{(1)}[x_k]&=&\ds\frac{v_k}{y_k}\left(\left(L_1^{i-1}\right)_{k+1,k}- \left(L_1^{i}\right)_{k+1,k+1}\right),\\
  \X_i^{(2)}[x_k]&=&\ds\frac{v_k}{y_k}\left(\left(L_2^{i-1}\right)_{k,k+1}- \left(L_2^{i}\right)_{k,k}\right).
\end{eqn}%
According to (\ref{one_vec}), the first equation amounts to 
\begin{equation}\label{tri}
  y_k\inn{L_1^{i-1}}{\pp{L_1}{y_k}}=\left(L_1^i\right)_{k+1,k+1}-\left(L_1^{i-1}\right)_{k+1,k}, 
\end{equation}
where we recall that $\inn{A}{B}=\Trace AB$. The proof of (\ref{tri}) follows immediately by writing
$(L_1^i)_{k+1,k+1}$ as $(L_1^{i-1}L_1)_{k+1,k+1}$, and the expression (\ref{lax_mats}) for the entries of
$L_1$. For the second equation in (\ref{to_pr}) the proof is similar.

Notice that (\ref{G_eq}) implies that the time-dependent polynomials $\G_k(t)$ and $\Gt_k(t)$ are given by
\begin{eqnarray*}
    \G_k(t)&=&\V^{u(t)}[x_k](t)+kx_k(t),\\
    \Gt_k(t)&=&-\V^{u(t)}[y_k](t)+ky_k(t),
\end{eqnarray*}
where $\V^{u(t)}$ can, in view of (\ref{u_def}) be written as
\begin{equation*}
  \V^{u(t)}=t(\X_1^{(1)}+\X_1^{(2)})+\V^u.
\end{equation*}%
Since the vector field $\p/\p t$ commutes with all the Hamiltonian vector fields $\X_i^{(1)}$ and $\X_i^{(2)}$, it
follows from these equations and (\ref{one_vec_bis}) that
{\small\begin{eqnarray*}
\dot\G_k(t)
  &=&\X_1^{(1)}[x_k](t)+\X_1^{(2)}[x_k](t)+\V^{u(t)}[\dot x_k](t)+k\dot x_k(t)\\
     &=&(k+1)\X_1^{(1)}[x_k](t)-(k-1)\X_1^{(2)}[x_k](t)+\V^{u(t)}\left[v_k(x_{k+1}-x_{k-1})\right](t)\\
     &=&(k+1)v_k(t)x_{k+1}(t)-(k-1)v_{k}(t)x_{k-1}(t)\\
     &&+v_k(t)\V^{u(t)}\left[x_{k+1}-x_{k-1}\right](t)-(x_{k+1}(t)-x_{k-1}(t))\V^{u(t)}\left[x_ky_k\right](t) \\
     &=&v_k(t)(\G_{k+1}(t)-\G_{k-1}(t))+(x_{k+1}(t)-x_{k-1}(t)) (x_k(t)\Gt_k(t)-y_k(t)\G_k(t)).
\end{eqnarray*}}%
%
This yields the first relation in (\ref{G_diff_eq}). The second equation is obtained by duality.

At points of $\M_u$ all $\G_k$ and $\Gt_k$ vanish so the right hand sides of (\ref{G_diff_eq}) vanish. The unique
solution to (\ref{G_diff_eq}) that corresponds to such initial data is the zero solution, $\G_k(t)=\Gt_k(t)=0$. As a
consequence, $\M_{u(t)}$ is a time-dependent invariant manifold for the first Toeplitz flow.
\qed
\end{proof}
%

%
\section{Painlev\'e analysis of the first Toeplitz flow}\label{laurent_sec}
In this section we will show that the first Toeplitz flow admits many families of formal Laurent solutions, a
property reminiscent of (finite-dimensional) algebraic completely integrable systems (see \cite{avmbook}). They
will be used in the subsequent chapters. We will first consider the self-dual case, which is easier, and then we
will consider the full Toeplitz lattice.\qquad

\subsection{The self-dual Toeplitz lattice}
Recall that the first vector field of the self-dual Toeplitz lattice is given by
\begin{equation}\label{self-dual_eqs}
  \dot x_k=(1-x_k^2)(x_{k+1}-x_{k-1}),\qquad k\in\Z,
\end{equation}
which we also write as $\dot x_k=v_k(x_{k+1}-x_{k-1})$, since $v_k:=1-x_k^2$, for $k\in\Z$.
\begin{proposition}\label{self_dual_laurent_prop}
  For any $n\in\Z$, the first vector field (\ref{self-dual_eqs}) of the self-dual Toeplitz lattice admits a formal
  Laurent solution $x(t)$, with only $x_n(t)$ having a pole, given by
\begin{eqnarray*}
  x_k(t)&=&\e\left(a_k+\check a_k(a_{k+1}-a_{k-1})t+\frac12\check a_k (a_{k-2}\check a_{k-1}+a_{k+2}\check a_{k+1}\right.\\ 
  &&\quad\left.-\,a_k((a_{k+1}-a_{k-1})^2+2-2a_{k-1}a_{k+1})+\kappa_k)t^2+ +O(t^3)\right),\\
           &&\qquad\qquad\qquad\qquad\qquad\qquad\qquad\qquad\qquad\qquad\quad \vert k-n\vert\geq 2,\\
           x_{n\pm1}(t)&=&\e\left(\mp1+4a_\pm t+4a_\pm(2a_{n\pm2}\mp(a_-+a_+))t^2 +O(t^3)\right),\\
           x_n(t)&=&-\frac{\e}{2t}\left(1+(a_+-a_-)t+\frac13((a_+-a_-)^2\right.\\
           &&\quad\left.+\,4(a_+a_{n+2}-a_-a_{n-2}+1-2a_+a_-))t^2+O(t^3)\right),
\end{eqnarray*}
  where $a_+,a_-$ and all $a_i$, with $i\in\Z\setminus\set{n-1,n,n+1}$ are arbitrary free parameters, $\check a_k$ is an
  abbreviation for $1-a_k^2$; also, $\e^2=1$ and $a_{n-1}=-a_{n+1}=1$. When $\vert k- n\vert >2$ then $\kappa_k=0$,
  while $\kappa_{n\pm2}=\mp4 a_\pm$.
\end{proposition}
\begin{proof}
We look for formal Laurent solutions $x(t)$ to (\ref{self-dual_eqs}) that have a simple pole for one of the
variables (only). To do this, we substitute $x_n(t)=x_n^{(0)}/t+O(1)$, with $x_n^{(0)}\neq0$, and
$x_j(t)=x_j^{(0)}+O(t),\, j\neq n$ into (\ref{self-dual_eqs}) for different values of $k$. Taking $k=n\pm1$ we find
that $\left(x_{n\pm1}^{(0)}\right)^2=1$, in both cases because $1-x_k^2(t)$ needs to cancel the pole coming from
$x_n(t)$. Given this, (\ref{self-dual_eqs}) with $k=n$ is given by
\begin{equation*} 
  -\frac{x_n^{(0)}}{t^2}+O(1)= -\frac{\left(x_n^{(0)}\right)^2}{t^2}(x_{n+1}^{(0)}-x_{n-1}^{(0)})+O(t^{-1}).
\end{equation*}%
Since $x_n^{(0)}\neq0$, we deduce from it on the one hand that $x^{(0)}_{n+1}$ and $x^{(0)}_{n-1}$ have opposite signs,
so that $x^{(0)}_{n+1}=-x^{(0)}_{n-1}$ and that $x_n^{(0)}=1/(2x_{n+1}^{(0)})$. It follows that
$x_{n\pm1}(t)=\mp\e+O(t)$ and $x_n(t)=-\e/(2t)+O(1)$, where $\e^2=1$.  For $\vert k-n\vert\geq2$, the coefficient in
$t^{-1}$ of (\ref{self-dual_eqs}) does not impose any condition on the constant coefficient of $x_k(t)$, which is
therefore a free parameter, which we denote as $\e a_k$.

\smallskip

Having determined the first term of the series we suppose that 
\begin{eqnarray*}
  x_k(t)&=&\e\left(a_k+\sum_{i=1}^rx_k^{(i)}t^i+x_k^{(r+1)}t^{r+1}\right), \quad\vert k-n\vert\geq2,\\
            x_{n\pm1}(t)&=&\e\left(\mp1+\sum_{i=1}^rx_{n\pm1}^{(i)}t^i+ x_{n\pm1}^{(r+1)}t^{r+1}\right),\\
            x_n(t)&=&-\frac{\e}{2t}\left(1+\sum_{i=1}^rx_n^{(i)}t^i+ x_n^{(r+1)}t^{r+1}\right),
\end{eqnarray*}
where all coefficients $x_k^{(i)}$, with $i\leq r$ have been determined. We show that (\ref{self-dual_eqs}) then yields
linear relations on the coefficients $x_k^{(r+1)}$. To see that, pick the coefficient in $t^{r}$ in
(\ref{self-dual_eqs}) when $k\neq n$, while taking the coefficient in $t^{r-1}$ when $k=n$. This yields the following
relations, where ``known'' means coefficients $x_k^{(i)}$, with $i\leq r$:
\begin{eqnarray}\label{sd_lau_rec}
  \vert k-n\vert\geq2&:&\e(r+1)x_k^{(r+1)}=\hbox{known},\nonumber\\
  k=n\pm1&:&\e rx_{n\pm1}^{(r+1)}=\hbox{known},\\   
  k=n&:&-\frac{\e}2(r+2)x_n^{(r+1)}=-\frac{\e}4(x_{n+1}^{(r+1)}-x_{n-1}^{(r+1)})+\hbox{known}.\nonumber
\end{eqnarray}
This yields a linear system in the unknowns $x_k^{(r+1)}$, where $k\in\Z$, which has upper triangular form when
$x_n^{(r+1)}$ is put at the end. It uniquely determines the coefficients $x_{k}^{(r+1)}$, except when $k=n\pm1$ and
$r=0$: the corresponding equations both reduce then to $0=0$, so that $x_{n+1}^{(1)}$ and $x_{n-1}^{(1)}$ are also
free parameters; we denote them by $4a_\pm:=x_{n\pm1}^{(1)}$. Then the third equation in (\ref{sd_lau_rec}) implies
that $x_n^{(1)}=a_+-a_-$; also, the first equation is explicitly given by $\e
x_k^{(1)}=\e(1-a_k^2)(a_{k+1}-a_{k-1})$, for $\vert k-n\vert\geq2$. Since for $r>0$ we can solve uniquely for all
$x_k^{r+1}$, we get a formal Laurent solution depending on the free parameters, as indicated. The extra term that
is given in the proposition is easily verified.
\qed
\end{proof}
Notice that under the natural correspondence between the phase variables~$x_k$ (with $k\neq n$) and the free
parameters $a_k$ ($a_\pm$ in the case $k=n\pm1$) we have that the number of free parameters on which the
coefficients of the series depend, is one less than the number of phase variables, a property reminiscent of
principal balances for (finite-dimensional) algebraic completely integrable systems (see \cite[Chapter
6]{avmbook}). There are of course also formal Laurent solutions that depend on less free parameters (lower
balances), but these will not be used here.

\smallskip

For future reference we give the first few terms of the formal Laurent series of $v_k:=1-x_k^2$, which is easily
computed from the series given in Proposition~\ref{self_dual_laurent_prop},
\begin{eqn}{v_k_sd}
  v_k(t)&=&\check a_k-2a_k\check a_k(a_{k+1}-a_{k-1})t+O(t^2),\qquad \vert k-n\vert\geq 2,\\ 
  v_{n\pm1}(t)&=&\pm8a_\pm t+O(t^2),\\ 
  v_n(t)&=&\displaystyle-\frac1{4t^2}(1+2(a_+-a_-)t+O(t^2)).
\end{eqn}%
The displayed terms are the only ones that will be needed below.

\subsection{The full Toeplitz lattice}
We will now show that the full Toeplitz lattice also allows such formal Laurent solutions. To make the analogy with the
self-dual case transparent we will vectorize the variables and the equations, namely we introduce $z_k:=\left(
\begin{array} {c} x_k\\y_k\end{array}\right)$ and $c_k:=\left(\begin{array}{c} a_k\\b_k\end{array}\right)$, for
$k\in\Z$; the variables $a_k$ and $b_k$ will be the free parameters in the formal Laurent series. With these
notations the first Toeplitz vector field (\ref{toeplitz_vf}) becomes
\begin{equation}\label{toeplitz_vf_bis}
  \dot z_k=(1-x_ky_k)(z_{k+1}-z_{k-1}).
\end{equation}
\begin{proposition}\label{general_laurent_prop}
  For any $n\in\Z$, the vector field (\ref{toeplitz_vf_bis}) of the (general) Toeplitz lattice admits a formal Laurent
  solution $z(t)=\begin{matrix}{c} x(t)\\y(t)\end{matrix}$, such that only $x_n(t)$ and $y_n(t)$ have a (simple)
  pole. It is given by
\begin{eqnarray*}
  z_k(t)&=&c_k+\check c_k(c_{k+1}-c_{k-1})t+O(t^2),\qquad \vert k-n\vert\geq 2,\\
  z_{n\pm1}(t)&=&
    \begin{matrix}{c} 
       a_{n\pm1}+a_{n\pm1}a_\pm t\\
       1/a_{n\pm1}-a_\pm/a_{n\mp1}t
    \end{matrix}+O(t^2)\\
  z_n(t)&=&\frac{1}{(a_{n-1}-a_{n+1})t}
     \begin{matrix}{c} 
       a_{n-1}a_{n+1}(1+at)\\
       -1+\frac{a_{n+1}(a_++a)-a_{n-1}(a_-+a)}{a_{n+1}-a_{n-1}}t
     \end{matrix}+O(t),
\end{eqnarray*}
  where $a,\,a_\pm,\,a_{n\pm1}$ and all $c_i=\begin{matrix}{c}a_i\\ b_i\end{matrix}$, with
  $i\in\Z\setminus\set{n-1,n,n+1}$ are arbitrary free parameters, and where $c_{n\pm1}=\begin{matrix}{c}a_{n\pm1}\\
  1/a_{n\pm1}\end{matrix}$.  Precisely, the free parameters $a_{n\pm1}$ satisfy the condition
  $a_{n+1}a_{n-1}(a_{n+1}-a_{n-1})\neq0$. Also, $\check c_k=1-a_kb_k$. The parameters on which the next order term
  in the series $x(t)$ and $y(t)$ depend is given in Table \ref{dependence_table}.
\end{proposition}
\begin{remark}
In Section \ref{conf_general_sec} we will need some extra information on these formal Laurent series, namely that the
coefficient in $t^2$ of $z_k$, for $\vert k-n\vert \geq2$ depends in the following way on $c_{k+2}$,
\begin{equation}
  z_k^{(2)}=\frac12\check c_k\check c_{k+1}c_{k+2}+\tilde z_k^{(2)},
\end{equation}
where $\tilde z_k^{(2)}$ is independent of $a_{k+2}$ and of $b_{k+2}$.  In particular, $x_k^{(2)}$ depends linearly
on $a_{k+2}$ and is independent of $b_{k+2}$, while $y_k^{(2)}$ depends linearly on $b_{k+2}$ and is independent of
$a_{k+2}$. This easily follows from the given terms by considering the coefficient of $t$ in
(\ref{toeplitz_vf_bis}).
\end{remark}

\begin{table}[ht]
  \caption{We list on which free parameters the first few terms of the formal Laurent solutions depend. It is understood that
  we do not list again the parameters that appear already before, on the same line; for example, $x_n^{(1)}$ depends
  only on $a_{n+1},\,a_{n-1}$ and $a$. The last two lines correspond to the values $k$ for which $\vert k-n\vert
  >2$. For $k\neq n$, $x_k^{(i)}$ is the coefficient of $t^i$ in $x_k(t)$, while for $k=n$ it is the coefficient of
  $t^{i-1}$ in $x_n(t)$.
  \label{dependence_table}}
  \begin{center}
  \renewcommand{\arraystretch}{1.5}
  \begin{tabular}{|c|c|c|c|c|}
    \hline 
    &$x^{(0)},y^{(0)}$&$x^{(1)},y^{(1)}$&$x^{(2)},y^{(2)}$\\
    \hline
    $x_n$&$a_{n+1},a_{n-1}$&$a$&$a_-, a_+, a_{n+2}, b_{n+2},a_{n-2},b_{n-2}$\\ 
    $y_n$&$a_{n+1},a_{n-1}$&$a,a_+,a_-$&$a_{n+2}, b_{n+2},a_{n-2},b_{n-2}$\\
    $x_{n\pm1}$&$a_{n\pm1}$&$a_\pm$&$a_{n\pm2},b_{n\pm2},a_\mp,a,a_{n\mp1}$\\
    $y_{n\pm1}$&$a_{n\pm1}$&$a_{n\mp1},a_\pm$&$a_{n\pm2},b_{n\pm2},a_\mp,a$\\
    $x_{n\pm2}$&$a_{n\pm2}$&$a_{n\pm3},a_{n\pm1},b_{n\pm2}$&$a_{n\pm4},b_{n\pm3},a_\pm$\\
    $y_{n\pm2}$&$b_{n\pm2}$&$b_{n\pm3},b_{n\pm1},a_{n\pm2}$&$b_{n\pm4},a_{n\pm3},a_\pm,a_{n\mp1}$\\
    $x_{k}$&$a_{k}$&$a_{k+1},a_{k-1},b_{k}$&$a_{k+2},b_{k+1},a_{k-2},b_{k-1}$\\
    $y_{k}$&$b_{k}$&$b_{k+1},b_{k-1},a_{k}$&$b_{k+2},a_{k+1},b_{k-2},a_{k-1}$\\
    \hline
  \end{tabular}
  \end{center}
\end{table}

\begin{proof}
For fixed $n\in\Z$, we look for formal Laurent solutions 
$z(t)=\left(\begin{array}{c}x(t)\\y(t)\end{array}\right),$
 to (\ref{toeplitz_vf_bis}) where $x_n(t)$ or $y_n(t)$ have a simple pole, and where none of the other variables
$x_k(t)$ or $y_k(t)$ have a pole (in $t$).  Thus, we substitute $z_n(t)=z_n^{(0)}/t+O(1)$ and
$z_j(t)=z_j^{(0)}+O(t),\, j\neq n$ into (\ref{toeplitz_vf_bis}) for different values of $k$. For $k=n\pm1$ we find
that $x_{n\pm1}^{(0)}y_{n\pm1}^{(0)}=1$, because $1-x_{n\pm1}y_{n\pm1}$ needs to cancel the pole coming from $x_n$
or from $y_n$; we put $a_{n\pm1}:=x_{n\pm1}^{(0)}$, so that $y_{n\pm1}^{(0)}=1/a_{n\pm1}$. The parameters
$a_{n\pm1}$ are free, except that $a_{n+1}a_{n-1}\neq0$. Next, (\ref{toeplitz_vf_bis}) with $k=n$, yields
\begin{equation*}
\renewcommand{\arraystretch}{1.8}
  \left(
  \begin{array}{c}
    x_n^{(0)}\\ y_n^{(0)}
  \end{array}
  \right)
  =
  \left(
  \begin{array}{c}
    x_{n+1}^{(0)}-  x_{n-1}^{(0)}\\
    y_{n+1}^{(0)}-  y_{n-1}^{(0)}
  \end{array}
  \right)
  x_n^{(0)}y_n^{(0)}
\end{equation*}%
which shows on the one hand that $x_n^{(0)}$ and $y_n^{(0)}$ are both different from zero (since at least one of them is
supposed to be different from zero), so that also $a_{n+1}-a_{n-1}\neq0$.  On the other hand it shows that $x_n^{(0)}$
and $y_n^{(0)}$ are expressible in terms of $a_{n+1}$ and $a_{n-1}$ as
\begin{equation*}
  x_n^{(0)}=\frac{a_{n+1}a_{n-1}}{a_{n-1}-a_{n+1}},\qquad
  y_n^{(0)}=\frac{1}{a_{n+1}-a_{n-1}}.
\end{equation*}%
For $\vert k-n\vert\geq2$, the coefficient in $t^{-1}$ of (\ref{toeplitz_vf_bis}) does not impose any condition on the
constant coefficient of $z_k(t)$, yielding free parameters for the constant coefficients of $x_k$ and of $y_k$, with
$\vert k-n\vert>1$. We denote these free parameters by $c_k=\left(\begin{array}{c} a_k\\b_k
\end{array}\right)$.  Upon specialization, some of the formulas below may
contain $c_{n+1}$ or $c_{n-1}$; it is understood that these stand for 
\begin{equation*}
  c_{n\pm1}=
   \left(\begin{array}{c} a_{n\pm1}\\b_{n\pm1}\end{array}\right)
  =\left(\begin{array}{c} a_{n\pm1}\\1/a_{n\pm1}\end{array}\right).
\end{equation*}%
We can now proceed as in the second part of the proof
of Proposition \ref{self_dual_laurent_prop}, namely we suppose that
\begin{eqnarray*}
  z_k(t)&=&c_k+\sum_{i=1}^rz_k^{(i)}t^i+z_k^{(r+1)}t^{r+1},\\
  z_{n\pm1}(t)&=&\left(\begin{array}{c} a_{n\pm1}\\1/a_{n\pm1}\end{array}\right)+\sum_{i=1}^rz_{n\pm1}^{(i)}t^i+
          z_{n\pm1}^{(r+1)}t^{r+1},\\
  z_n(t)&=&\frac{1}{(a_{n-1}-a_{n+1})t}\left(\left(\begin{array}{c} a_{n-1}a_{n+1}\\-1\end{array}\right)
          +\sum_{i=1}^rz_n^{(i)}t^i+z_n^{(r+1)}t^{r+1}\right),
\end{eqnarray*}
where all coefficients $z_k^{(i)}$, with $i\leq r$ have been determined. On the coefficients $z_k^{(r+1)}$, $k\in\Z$, we
find linear relations by substituting the above series into (\ref{toeplitz_vf_bis}). For $k$ such that $\vert
n-k\vert>1$ it is clear that, as in the self-dual case, $z_k^{(r+1)}$ is linearly computed in terms of the known
coefficients, from the coefficient of $t^r$, when substituting the series in (\ref{toeplitz_vf_bis}). Therefore, let us
concentrate on what happens for $k\in\set{n-1,n,n+1}$. Taking $k=n\pm1$ in (\ref{toeplitz_vf_bis}) the coefficient of
$t^r$ yields
\begin{equation*}
  \renewcommand{\arraystretch}{1.8}
  (r+1)z_{n\pm1}^{(r+1)}=\pm
  \left(\frac{x_{n\pm1}^{(r+1)}}{a_{n\pm1}}+y_{n\pm1}^{(r+1)}a_{n\pm1}\right)
  \left(\begin{array}{c} \frac{a_{n-1}a_{n+1}}{a_{n-1}-a_{n+1}}\\
   \frac{-1}{a_{n-1}-a_{n+1}}\end{array}\right)  +\hbox{known},
\end{equation*}%
a linear equation in $x_{n\pm1}$ and $y_{n\pm1}$, which can be written in the compact form
\begin{equation*}
  \left(\L_\pm+(r+1)\Id\right)z_{n\pm1}^{(r+1)}=\hbox{known},
\end{equation*}%
where $\L_\pm$ is the matrix that governs the linear problem,
\begin{equation*}
  \L_\pm:=\pm\frac1{a_{n-1}-a_{n+1}}
  \begin{matrix}{cc}
    -a_{n\mp1}&-a_{n-1}a_{n+1}a_{n\pm1}\\
    {1/a_{n\pm1}}&a_{n\pm1}
  \end{matrix}.
\end{equation*}%
Since $\det(\L_\pm+(r+1)\Id)=r(r+1)$ this linear system admits a unique solution, except when $r=0$ (recall that
$r\geq0$). Before analyzing the case $r=0$ further, let us first consider what happens to (\ref{toeplitz_vf_bis}) in the
remaining case $k=n$. As in the self-dual case, we pick the coefficient of $t^{r-1}$ in (\ref{toeplitz_vf_bis}) to find
a linear system that can be written in the compact form
\begin{equation*}
  \left(\L_n+r\Id\right)z_{n}^{(r+1)}=\hbox{known},
\end{equation*}%
where the matrix $\L_n$ is given by
\begin{equation*}
  \L_n:=
  \begin{matrix}{cc}
    1&-a_{n+1}a_{n-1}\\
    -{1}/{(a_{n+1}a_{n-1})}&1
  \end{matrix}.
\end{equation*}%
Since $\det(\L_n+r\Id)=r(r+2)$ we have again that $z_n^{(r+1)}$ is determined uniquely, unless $r=0$. Thus, we are
done with $r\geq1$.

As we have seen, a free parameter may appear in $z_{n+1}^{(1)}$, in $z_{n-1}^{(1)}$ and in $z_n^{(1)}$, but one has to
check that the corresponding linear equations are consistent. Therefore we substitute
\begin{eqnarray}\label{subs1}
  z_k(t)&=&c_k+z_k^{(1)}t+O(t^2),\nonumber\\
  z_{n\pm1}(t)&=&\left(\begin{array}{c} a_{n\pm1}\\1/a_{n\pm1}\end{array}\right)+ z_{n\pm1}^{(1)}t+O(t^2),\\
  z_n(t)&=&\frac{1}{(a_{n-1}-a_{n+1})t}\left(\left(\begin{array}{c} a_{n-1}a_{n+1}\\-1\end{array}\right)
          +z_n^{(1)}t+O(t^2)\right),\nonumber
\end{eqnarray}
in (\ref{toeplitz_vf_bis}), which yields for $k=n\pm1$ and $t=0$ the homogeneous linear system
\begin{equation*}
  \renewcommand{\arraystretch}{1.8}
    \begin{matrix}{c}
      x_{n\pm1}^{(1)}\\
      y_{n\pm1}^{(1)}
    \end{matrix}
    =
    \pm\frac1{a_{n-1}-a_{n+1}}
    \left(\frac{x_{n\pm1}^{(1)}}{a_{n\pm1}}+y_{n\pm1}^{(1)}a_{n\pm1}\right)
    \begin{matrix}{c} 
      a_{n-1}a_{n+1}\\ -1
    \end{matrix},
\end{equation*}%
which is equivalent to 
\begin{equation}\label{par_rel}
  x_{n\pm1}^{(1)}+a_{n-1}a_{n+1}y_{n\pm1}^{(1)}=0.  
\end{equation}
Thus, upon setting $x_{n\pm1}^{(1)}=a_{n\pm1}a_{\pm}$, where $a_+$ and $a_-$ are free parameters, we have that
$y_{n\pm1}^{(1)}=-a_\pm/a_{n\mp1}=-a_\pm b_{n\mp1}$. Similarly, for $k=n$ the substitution of the series
(\ref{subs1}) in (\ref{toeplitz_vf_bis}) yields at the level $t^{-1}$:
\begin{eqnarray*}
  \frac{a_{n-1}a_{n+1}}{a_{n-1}-a_{n+1}}(x_{n+1}^{(1)}-x_{n-1}^{(1)}) -x_n^{(1)}+a_{n-1}a_{n+1}y_n^{(1)}=0,\\
  \frac{a_{n-1}a_{n+1}}{a_{n-1}-a_{n+1}}(y_{n+1}^{(1)}-y_{n-1}^{(1)}) -y_n^{(1)}+\frac{x_n^{(1)}}{a_{n-1}a_{n+1}}=0.
\end{eqnarray*}
These equation are proportional, in view of (\ref{par_rel}). Thus we have
\begin{eqnarray*}
  x_n^{(1)}&=&a_{n+1}a_{n-1}a,\\
  y_{n}^{(1)}&=&a+\frac{a_{n+1}a_+-a_{n-1}a_-}{a_{n+1}-a_{n-1}},
\end{eqnarray*}
where $a$ is a free parameter.

\smallskip

The first two terms in the series lead at once to the second and third columns of Table \ref{dependence_table}. In
order to obtain the last column it suffices to list on which parameters the linear term (resp.\ the constant term)
in the right hand side of $(1-x_k(t)y_k(t))(z_{k+1}(t)-z_{k-1}(t))$ depends, when $k\neq n$ (resp.\ when
$k=n$). The two leading terms of $x(t)$ and $y(t)$ that we computed suffice for doing this.
\qed
\end{proof}

It is easily verified that the involution $\s$, that permutes $x_k$ and $y_k$ extends naturally to an involution on the
free parameters, given by
\begin{eqn}[1.8]{s_ext}
  &\s(a_k)=b_k,\ \s(a_{n\pm1})=1/a_{n\pm1},\ \s(a_\pm)=-a_\pm a_{n\pm1}/a_{n\mp1}, \\
  &\s(a)=\ds-a-\frac{a_{n+1}a_+-a_{n-1}a_-}{a_{n+1}-a_{n-1}}.
\end{eqn}%

Notice that, altogether, we have besides the free parameters $a_{k}, b_k$, for $\vert k-n\vert>1$, which naturally
correspond to the variables $x_k$ and $y_k$, five extra free parameters $a_{n\pm1},\,a_\pm$ and $a$, that correspond to
the remaining six variables $x_{n\pm1},\,y_{n\pm1}$ and $x_n,\,y_n$, which again yields that the number of free
parameters, plus time,  is equal to the number of phase variables. This count will be important, and rigorous, when we restrict
these formal Laurent solutions to certain finite-dimensional submanifolds.

%
\section{Tangency to $\M$}
\label{tangent_sec}
We have seen that the polynomials $\G_k$ and $\Gt_k$, which define an invariant manifold for the first Toeplitz
flow, satisfy a non-autonomous system of linear differential equations, where the time-dependence is defined by the
latter flow. In a (finite-dimensional) manifold setting, if such differential equations have coefficients that
depend smoothly on time, solutions (integral curves) that start out on the invariant manifold will stay on it, by
the uniqueness of solutions to differential equations with smooth coefficients and given initial conditions. In the
case that we deal with the situation is quite a bit different, because the coefficients develop poles in~$t$, for
$t=0$, and of course the solutions are only formal Laurent series. As it turns out, the conditions that assure that
the formal Laurent solutions ``stay on the invariant manifold'' are similar to those in the smooth case for the
self-dual Toeplitz lattice, but are different in an essential way for the general Toeplitz lattice.%
 
\subsection{Tangency in the self-dual case}
We start out with the case of the self-dual Toeplitz lattice. 
\begin{proposition}\label{self-dual_tan_prop}
  Let $x(t)$ denote the formal Laurent solution that is given by Proposition \ref{self_dual_laurent_prop}, and let
  $\G(t):=\G(x(t);u(t))$, where we recall that $u(t)=(u_1+t,u_2,\dots,u_N)$. Then, as formal series in $t$,
  \begin{eqn}{gn_selfdual}
    \G_k(t)&=&\G_k^{(0)}+O(t), \qquad\qquad k\in\Z\setminus\set{n},\\
    \G_n(t)&=&\displaystyle\frac{1}{4t}(\G_{n+1}^{(0)}-\G_{n-1}^{(0)})+\G_n^{(0)}+O(t).
  \end{eqn}%
  Moreover, $\G_k(t)=0$ as a formal series in $t$, for all $k\in\Z$, as soon as $x(t)$ is such that
\begin{equation*}
  \G_k^{(0)}=0, \hbox{ for all $k\in\Z$}.  
\end{equation*}%
\end{proposition}
\begin{proof}
%
According to (\ref{high_selfdual}), $\G_k(x;u)$ involves only the variables $x_l$ with $\vert l-k\vert\leq N$ ($2N+1$
step relation). Since only $x_n(t)$ has a pole, $\G_k(t)=O(1)$ as soon as $\G_k$ does not contain $x_n$, i.e., if $\vert
n-k\vert>N$. But notice that (\ref{G_diff_eq_selfdual}) implies
\begin{equation*}
  \G_{n-N}=\frac{\dot\G_{n-N-1}}{v_{n-N-1}}+\G_{n-N-2},
\end{equation*}%
so that $\G_{n-N}(t)=O(1)$, as the leading term $\check a_{n-N-1}=1-a_{n-N-1}^2$ of $v_{n-N-1}(t)$ is non-zero (recall
that $a_{n-N-1}$ is a free parameter). This argument can be repeated to yield $\G_k(t)=O(1)$ for all $k<n$, and
similarly it is shown that $\G_k(t)=O(1)$ for all $k>n$. Since $\G_n(t)$ satisfies the differential equation
(\ref{G_diff_eq_selfdual}), for $k=n$, we have in view of (\ref{v_k_sd}) that
\begin{equation*}
  \frac{d\G_n}{dt}(t)=v_n(t)(\G_{n+1}(t)-\G_{n-1}(t))
   =-\frac1{4t^2}(\G_{n+1}^{(0)}-\G_{n-1}^{(0)})+O(1),
\end{equation*}%
which leads upon integration to (\ref{gn_selfdual}). 

Suppose now that $x(t)$ is such that $\G_k^{(0)}=0$ for all $k\in\Z$. In view of the first part of the proof, we
have that $\G_k(t)=O(t)$ for all $k\in\Z$. We show that this implies that $\G_k(t)=0$ as a formal series in $t$,
for all $k\in\Z$. We do this by induction on $r\in\N^*$: assuming that $\G_k(t)=O(t^r)$ for $k\in\Z$ we show that
$\G_k(t)=O(t^{r+1})$ for $k\in\Z$. Notice that in the case $r=1$ the assumption holds. For $k\notin\set{n-1,n,n+1}$
the right hand side of (\ref{G_diff_eq_selfdual}) is $O(t^r)$, by (\ref{v_k_sd}) and by the assumption, so that
$\dot\G_k(t)=O(t^r)$, hence $\G_k(t)=O(t^{r+1})$, by integration. For $k=n\pm1$ we have from (\ref{v_k_sd}) that
$v_{n\pm1}(t)=O(t)$, so that (\ref{G_diff_eq_selfdual}) yields for $k=n\pm1$ that $\dot\G_{n\pm1}(t)=O(t^{r+1})$,
i.e., $\G_{n\pm1}(t)=O(t^{r+2})$. For $k=n$ we have that $v_n(t)=1-x_n^2(t)$ has a double pole, but since we have
just shown that $\G_{n+1}(t)-\G_{n-1}(t)=O(t^{r+2})$ the differential equation (\ref{G_diff_eq_selfdual}) for $k=n$
leads to $\dot\G_n(t)=O(t^{r})$ and we conclude that $\G_n(t)=O(t^{r+1})$, as was to be shown.
\qed
\end{proof}

\subsection{Tangency in the general case}
For the full Toeplitz lattice the tangency condition is rather similar, yet is different in some detail that will turn
out to be crucial in the next section. We recall that the differential equations that are satisfied by the polynomials
$\G_k$ and $\Gt_k$ are given by
\begin{eqn}{G_diff_eq_k}
  \dot\G_k&=&v_k(\G_{k+1}-\G_{k-1})+(x_{k+1}-x_{k-1})(x_k\Gt_k-y_k\G_k),\\
  \dot{\Gt}_k&=&v_k(\Gt_{k+1}-\Gt_{k-1})-(y_{k+1}-y_{k-1})(x_k\Gt_k-y_k\G_k).
\end{eqn}%
\begin{proposition}\label{tang_prop}
  Let $(x(t),y(t))$ denote the formal Laurent solution that is given by Proposition \ref{general_laurent_prop}, and
  let $\G(t):=\G(x(t),y(t);u(t))$, where $u(t)$ is given by (\ref{u_def}). Then, as a formal series in $t$,
  $\G_k(t)=\G_k^{(0)}+O(t)$ and $\Gt_k(t)=\Gt_k^{(0)}+O(t)$ for $k\in\Z\setminus\set{n}$. Also
  \begin{eqn}[2]{gn_general}
    \G_n(t)&=&\displaystyle\frac{a_{n+1}^2}{a_-(a_{n-1}-a_{n+1})^2t^2}
          \left(\G_{n-1}^{(0)}-a_{n-1}^2\Gt_{n-1}^{(0)}\right)+
             \frac1t\G_n^{(-1)}+O(1),\\
     \Gt_n(t)&=&\displaystyle
           \frac{a_{n+1}a_{n-1}}{a_-(a_{n-1}-a_{n+1})^2t^2}
         \left(\G_{n-1}^{(0)}/a_{n-1}^2-\Gt_{n-1}^{(0)}\right)
             +\frac1t\Gt_n^{(-1)}+O(1),
  \end{eqn}%
  where $\G_n^{(-1)}$ and $\Gt_n^{(-1)}$ are both linear combinations of $\G_{n\pm1}^{(0)}$ and $\Gt_{n\pm1}^{(0)}$ (for
  the explicit formula, see (\ref{explicit})); moreover, the latter coefficients are related in the following way:
\begin{equation}\label{2way}
  a_-\left(\Gt_{n+1}^{(0)}-\frac1{a_{n+1}^2}\G_{n+1}^{(0)}\right)=
  a_+\left(\frac1{a_{n-1}^2}\G_{n-1}^{(0)}-\Gt_{n-1}^{(0)}\right).
\end{equation}
\end{proposition}
\begin{proof}
As in the self-dual case, the polynomials $\G_k(x;u)$ and $\Gt_k(x;u)$ define $2N+1$ step relations, so they depend only
on the variables $x_l$ and $y_l$ with $\vert l-k\vert\leq N$. Only $x_n(t)$ and $y_n(t)$ have a pole, so that
$\G_k(t)=O(1)$ and $\Gt_k(t)=O(1)$ for $\vert n-k\vert>N$. Writing (\ref{G_diff_eq_k}) for $k\to k-1$ as
\begin{eqn}[2]{GK_sol}
  \G_{k}&=&\displaystyle
          \frac1{v_{k-1}}\left(\dot\G_{k-1}-(x_{k}-x_{k-2})(x_{k-1}
          \Gt_{k-1}-y_{k-1}\G_{k-1})\right)+\G_{k-2},\\
  \Gt_{k}&=&\displaystyle
           \frac1{v_{k-1}}\left(\dot{\Gt}_{k-1}+(y_{k}-y_{k-2})(x_{k-1}
           \Gt_{k-1}-y_{k-1}\G_{k-1})\right)+\Gt_{k-2},
\end{eqn}%
and taking as consecutive values $k:=n-N,\dots,n-1$ in (\ref{GK_sol}) we find that $\G_k(t)=O(1)$ and $\Gt_k(t)=O(1)$
for all $k\leq n-1$, since $v_k(t)$ does not vanish for $t=0$ when $k\neq n\pm1$. Similarly $\G_k(t)=O(1)$ and
$\Gt_k(t)=O(1)$ when $k\geq n+1$. So we have that $\G_k(t)=O(1)$ and $\Gt_k(t)=O(1)$ when $k\neq n$ and we are left with
the case $k=n$.

\smallskip

In order to deal with the case $k=n$ we write (\ref{G_diff_eq_k}) as an equation for $\G_n$ and $\Gt_n$ in two different
ways:
\begin{eqn}[2]{GK_sol_n}
  \G_{n}&=&\displaystyle\mp \frac1{v_{n\pm1}}\left(\dot\G_{n\pm1}\pm(x_{n}-x_{n\pm2})(x_{n\pm1}
          \Gt_{n\pm1}-y_{n\pm1}\G_{n\pm1})\right)+\G_{n\pm2},\\ 
  \Gt_{n}&=&\displaystyle\mp\frac1{v_{n\pm1}}\left(\dot{\Gt}_{n\pm1}\mp(y_{n}-y_{n\pm2})(x_{n\pm1}
          \Gt_{n\pm1}-y_{n\pm1}\G_{n\pm1})\right)+\Gt_{n\pm2}.
\end{eqn}%
Either of them implies that $\G_n(t)=O(t^{-2})$ and that $\Gt_n(t)=O(t^{-2})$, so we write
\begin{equation*}
  \G_n(t)=\frac1{t^2}\left(\G_n^{(-2)}+\G_n^{(-1)}t+\G_n^{(0)}t^2+O(t^3)\right),
\end{equation*}%
and similarly for $\Gt_n(t)$. In fact, as $v_{n+1}(t)$ and $v_{n-1}(t)$ have a simple zero, while $x_n(t)$ and $y_n(t)$
have a simple pole, the coefficient of $t^{-2}$ in (\ref{GK_sol_n}), leads to the following linear equations
\begin{eqn}[1.8]{gn-2}
    \G_n^{(-2)}&=&\displaystyle-{x_n^{(0)}} \left(x_{n\pm1}^{(0)}\Gt_{n\pm1}^{(0)}-y_{n\pm1}^{(0)}
       \G_{n\pm1}^{(0)}\right)/{v_{n\pm1}^{(0)}},\\ 
    \Gt_n^{(-2)}&=&\displaystyle-\G_n^{(-2)}y_n^{(0)}/x_n^{(0)},
\end{eqn}%
where we have written $v_{n\pm1}(t)=v_{n\pm1}^{(0)}t+O(t^2)$, so that
\begin{equation}\label{vlead}
  v_{n\pm1}^{(0)}=\pm a_\pm\frac{a_{n+1}-a_{n-1}}{a_{n\mp1}}.
\end{equation}
It suffices now to substitue $x_{n\pm1}^{(0)}=a_{n\pm1}=1/y_{n\pm1}^{(0)}$ and $x_n^{(0)}=a_{n-1}a_{n+1}/$
$(a_{n-1}-a_{n+1})=-a_{n-1}a_{n+1}y_n^{(0)}$ in (\ref{gn-2}) to find the coefficient of $t^{-2}$ in
(\ref{gn_general}). Actually, the latter corresponds to taking the lower sign; equating the two expressions for
$\G_n^{(-2)}$ in (\ref{gn-2}) that correspond to the two signs leads to (\ref{2way}); notice that this is also the
expression that is obtained from the two expressions of $\Gt_n^{(-2)}$ in (\ref{gn-2}).

\smallskip

It remains to compute $\G_n^{(-1)}$ and $\Gt_n^{(-1)}$, which can be done from the coefficient of $t^{-2}$ in
$\dot\G_n(t)$ and in $\dot{\Gt}_n(t)$, computed from their differential equations
\begin{eqn}{G_diff_eq_n}
  \dot\G_n&=&v_n(\G_{n+1}-\G_{n-1})+(x_{n+1}-x_{n-1})(x_n\Gt_n-y_n\G_n),\\
  \dot{\Gt}_n&=&v_n(\Gt_{n+1}-\Gt_{n-1})-(y_{n+1}-y_{n-1})(x_n\Gt_n-y_n\G_n).
\end{eqn}%
Since $v_n(t)$ has a double pole, while $\G_{n\pm1}(t)$ and $\Gt_{n\pm1}(t)$ have no pole, the contribution of the first
term to the coefficient in $t^2$ will be linear in $\G_{n\pm1}^{(0)}$ and in $\Gt_{n\pm1}^{(0)}$. Since $x_n(t)$ and
$y_n(t)$ have a simple pole, while $\G_n(t)$ and $\Gt_n(t)$ have a double pole, the contribution of the second term will
yield a linear combination of on the one hand $\G_n^{(-2)}$ and $\Gt_n^{(-2)}$ which, as we have seen, are themselves
linear combinations of $\G_{n\pm1}^{(0)}$ and in $\Gt_{n\pm1}^{(0)}$; on the other hand, $\G_n^{(-1)}$ and
$\Gt_n^{(-1)}$, which are the unknowns. Explicitly, this linear system is given by
\begin{eqnarray}\label{explicit}
  \begin{matrix}{c}
    a_{n+1}a_{n-1}\Gt_n^{(-1)}\\
    1/(a_{n+1}a_{n-1})\G_n^{(-1)}    
  \end{matrix}
  =\frac{a_{n-1}a_{n+1}}{(a_{n+1}-a_{n-1})^2}
  \begin{matrix}{c}
    \G_{n+1}^{(0)}-\G_{n-1}^{(0)}\\
    \Gt_{n+1}^{(0)}-\Gt_{n-1}^{(0)}
  \end{matrix}-\nonumber\\
    \begin{matrix}{c}
      1\\
      1/a_{n+1}a_{n-1}
  \end{matrix}
  \left(\G_n^{(-2)}\s(a)+\Gt_n^{(-2)}aa_{n+1}a_{n-1}\right).\nonumber\\
\end{eqnarray}
Since $\G_n^{(-2)}$ and $\Gt_n^{(-2)}$ are linear combinations of $\G_{n\pm1}^{(0)}$ and $\Gt_{n\pm1}^{(0)}$ it follows
that each of $\G_n^{(-1)}$ and $\Gt_n^{(-1)}$ is a linear combination of $\G_{n\pm1}^{(0)}$ and $\Gt_{n\pm1}^{(0)}$, as
we asserted.
\qed
\end{proof}
%
\begin{proposition}\label{tang_gen_prop}
  Suppose that $(x(t),y(t))$ is a formal Laurent solution of the first vector field of the Toeplitz lattice, such
  that $\G_k(t)=O(t)$ and $\Gt_k(t)=O(t)$ for all $k$ with $k\neq n+1$, and such that, as formal Laurent solutions
  in $t$, $\G_{n-1}(t)=O(t^2)$ and $\G_{n+1}(t)=O(t)$. Then, as formal Laurent series, $\G_k(t)=0=\Gt_k(t)$ for
  all $k\in\Z$.
\end{proposition}
\begin{proof}
According to (\ref{2way}), the hypothesis imply that $\Gt_{n+1}(t)=O(t)$. In view of Proposition \ref{tang_prop},
we have that $\G_k(t)=O(t)$ and $\Gt_k(t)=O(t)$ for every $k\in\Z$. We will now proceed by induction on $r\in\N^*$,
but in a different way than in the self-dual case: assuming that $\G_k(t)=O(t^r)$ and $\Gt_k(t)=O(t^r)$ for $k\neq
n\pm1$, as well as $\G_{n\pm1}(t)=O(t^{r+1})$ and $\Gt_{n\pm1}(t)=O(t^{r+1})$, we show that $\G_k(t)=O(t^{r+1})$
and $\Gt_k(t)=O(t^{r+1})$ for $k\neq n\pm1$, as well as $\G_{n\pm1}(t)=O(t^{r+2})$ and
$\Gt_{n\pm1}(t)=O(t^{r+2})$. Notice that the $r=1$ induction assumption needs to be shown at the end of the proof,
as only part of it is in the actual hypothesis of the theorem.

\smallskip

For $k$ such that $\vert k-n\vert\geq2$ the differential equations (\ref{G_diff_eq_k}) yield that
$\dot\G_k(t)=O(t^r)$ and $\dot{\Gt}_k(t)=O(t^r)$, so that $\G_k(t)=O(t^{r+1})$ and $\Gt_{k}(t)=O(t^{r+1})$, by
integration. So we are left with $k\in\set{n-1,n,n+1}$. Let us write
\begin{equation*}
  \renewcommand{\arraystretch}{1.5}
  \begin{array}{rclrcl}
    \G_{n}&=&\gamma_{n}t^{r}+O(t^{r+1}),
    &\quad\Gt_{n}&=&\tilde\gamma_{n}t^{r}+O(t^{r+1}),\\
    \G_{k}&=&\gamma_{k}t^{r+1}+O(t^{r+2}),
    &\Gt_{k}&=&\tilde\gamma_{k}t^{r+1}+O(t^{r+2}),\quad k\neq n,
  \end{array}
\end{equation*}%
which we substitute in 
\begin{eqn}{my_rec_1}
  \dot\G_{n\pm1}&=&\mp v_{n\pm1}(\G_{n}-\G_{n\pm2})\pm(x_{n\pm2}-x_{n}) (x_{n\pm1}\Gt_{n\pm1}-y_{n\pm1}\G_{n\pm1}),\\
       \dot{\Gt}_{n\pm1}&=&\mp v_{n\pm1}(\Gt_{n}-\Gt_{n\pm2})\mp(y_{n\pm2}-y_{n}) (x_{n\pm1}\Gt_{n\pm1}-
       y_{n\pm1}\G_{n\pm1}).
\end{eqn}%
Remembering that $v_{n\pm1}(t)=O(t)$ we pick the coefficient of $t^r$ in (\ref{my_rec_1}), which leads to the
following linear system,
\begin{eqn}{rec_lin}
  (r+1)\gamma_{n\pm1}&=&\mp\frac{a_{n-1}a_{n+1}}{a_{n-1}-a_{n+1}}
   \left(a_{n\pm1}\tilde\gamma_{n\pm1}-\frac1{a_{n\pm1}}\gamma_{n\pm1}\right),\\
   (r+1)\tilde\gamma_{n\pm1}&=&\mp\frac1{a_{n-1}-a_{n+1}}
   \left(a_{n\pm1}\tilde\gamma_{n\pm1}-\frac1{a_{n\pm1}}\gamma_{n\pm1}\right).
\end{eqn}%
Since
\begin{equation*}
  \renewcommand{\arraystretch}{1.7}
  \begin{detmat}{cc}
   r+1\mp\frac{a_{n\mp1}}{a_{n-1}-a_{n+1}}&\pm\frac{a_{n-1}a_{n+1}a_{n\pm1}}{a_{n-1}-a_{n+1}}\\
  \mp\frac1{(a_{n-1}-a_{n+1})a_{n\pm1}}&\ r+1\pm\frac{a_{n\pm1}}{a_{n-1}-a_{n+1}}
  \end{detmat}
  =(r+1)^2-(r+1)=r(r+1),
\end{equation*}%
it follows, since $r\geq1$, that $\gamma_{n\pm1}=\tilde\gamma_{n\pm1}=0$, and hence that $\G_{n\pm1}(t)=O(t^{r+2})$
and $\Gt_{n\pm1}(t)=O(t^{r+2})$. It follows that, if we substitute the series in
\begin{eqn}{Gn_again}
  \dot\G_n&=&v_n(\G_{n+1}-\G_{n-1})+(x_{n+1}-x_{n-1})(x_n\Gt_n-y_n\G_n),\\
  \dot{\Gt}_n&=&v_n(\Gt_{n+1}-\Gt_{n-1})-(y_{n+1}-y_{n-1})(x_n\Gt_n-y_n\G_n),
\end{eqn}%
then the  coefficient of $t^{r-1}$ is simply given by
\begin{eqnarray*}
  r\gamma_n&=&-(a_{n-1}a_{n+1}\tilde\gamma_n+\gamma_n),\\
  r\tilde\gamma_n&=&-\frac1{a_{n-1}a_{n+1}}(a_{n-1}a_{n+1}\tilde\gamma_n+\gamma_n).
\end{eqnarray*}
Since
\begin{equation*}
  \det  
  \renewcommand{\arraystretch}{1.7}
  \begin{matrix}{cc}
    r+1&a_{n-1}a_{n+1}\\
    \frac1{a_{n+1}a_{n-1}}&r+1
  \end{matrix}
  =(r+1)^2-1\neq0,
\end{equation*}%
we have that $\gamma_n=\tilde\gamma_n=0$, so that $\G_n(t)=O(t^{r+1})$ and $\Gt_n(t)=O(t^{r+1})$, as was to be shown.

\smallskip

We finally check that our assumptions imply that for $r=1$ the induction hypothesis is valid. According to
Proposition \ref{tang_prop}, we have that $\G(t)=O(t)$ and $\Gt(t)=O(t)$. Let us write
$\G_{n\pm1}=\gamma_{n\pm1}t+O(t^2)$ and $\Gt_{n\pm1}=\tilde\gamma_{n\pm1}t+O(t^2)$. Then we need to show that
$\gamma_{n\pm1}=\tilde\gamma_{n\pm1}=0$. From (\ref{rec_lin}), which is also valid for $r=0$, we conclude that
$\gamma_{n\pm1}=a_{n-1}a_{n+1}\tilde\gamma_{n\pm1}$. It was assumed that $\G_{n-1}(t)=O(t^2)$, i.e., that
$\gamma_{n-1}=0$, so that we can conclude that $\tilde\gamma_{n-1}=0$. In order to obtain a second relation between
$\gamma_{n+1}$ and $\tilde\gamma_{n+1}$ we consider the residue in the first\footnote{Taking the second equation
would lead to the same result.}  equation in (\ref{Gn_again}), which reduces to
$0=a_{n-1}a_{n+1}\gamma_{n+1}/(a_{n-1}-a_{n+1})^2$, since $\G_n^{(0)}=\Gt_n^{(0)}=0$. Thus,
$\gamma_{n+1}=\tilde\gamma_{n+1}=0$, as was to be shown.
\qed
\end{proof}
%
  
%
\section{Restricting the formal Laurent solutions: the self-dual case}
\label{conf_self-dual_sec}
We have seen conditions on $\G(t)=\G(x(t);u(t))$ that guarantee that solutions $x(t)$ to the self-dual Toeplitz
lattice that start out in the invariant manifold $\M_{u(t)}$ stay in it, formally speaking. In this section we show
how these conditions can be translated into conditions on the formal Laurent solution $x(t)$ to the first vector
field of the self-dual Toeplitz lattice.
\subsection{Structure of the polynomials $\G_k$}
The polynomials $\G_k$, which define the invariant manifolds $\M$ depend on the variable $x_n$ in a special way,
that we will analyze by using the fact that $\G_k$ remains pole free (for $k\neq n$) when the formal Laurent series
$x(t)$ are substituted in them, as we have seen in Proposition \ref{self-dual_tan_prop}. Let us denote by $\A$ the
algebra of polynomials in all variables $x_k$, where $k\in\Z$ and by $\A_n$ the subalgebra of those polynomials
that are independent of $x_n$. Also, let us denote by $\A_n'$ the subalgebra of $\A$ that consists of those
elements that can be written as polynomials in $w_1, w_2$ and $x_k$, with $k\neq n$, where
\begin{equation}\label{w_def}
  w_1:=x_n(x_{n+1}+x_{n-1}),\ w_2:=x_n(1+x_{n+1}x_{n-1}).
\end{equation}
Thus, elements of $\A_n'$ may depend only on $x_n$ through $w_1$ and $w_2$. For future use, we give the first few
terms of the formal Laurent series of the generators of $\A_n'$, as obtained by substituting the series from Proposition
\ref{self_dual_laurent_prop} in (\ref{w_def}):
  \begin{eqn}{ws}
    w_1(t)&=&
         -2(a_++a_-+2(a_+a_{n+2}+a_-a_{n-2})t+O(t^2)),\\
    w_2(t)&=&
         -2\e(a_+-a_-+2(a_+a_{n+2}-a_-a_{n-2})t+O(t^2)),\\
    x_k(t)&=&\e (a_k+(1-a_k^2)(a_{k+1}-a_{k-1})t+O(t^2)),\quad k\neq n.
  \end{eqn}%
It follows that $G(x(t))=O(1)$, for any $G\in\A_n'$. Notice that the polynomials $w_\pm:=v_{n\pm1}x_n$, which both
have the property $w_\pm(t)=O(1)$, belong to $\A_n'$, since
\begin{equation}\label{wpmin}
  (1-x_{n\pm1}^2)x_n=w_2-x_{n\pm1}w_1.
\end{equation}
The following proposition generalizes this statement.
\begin{proposition}\label{pol_char_lemma}
  For $G\in\A$, let $G(t):=G(x(t))$, where $x(t)$ is the formal Laurent solution to the first vector field of the
  self-dual Toeplitz lattice, constructed in Proposition \ref{self_dual_laurent_prop}. If $G(t)=O(1)$ then
  $G\in\A_n'$, i.e., $G$ is a polynomial in
  \begin{equation*}
     x_n(x_{n+1}-x_{n-1}),\,x_n(1+x_{n+1}x_{n-1}),\textrm{ and } x_k\ (k\neq n).
  \end{equation*}%
\end{proposition}
\begin{proof}
We suppose that $G\in\A$ is such that $G(t)=O(1)$, where $G(t):=G(x(t))$. We write $G$ as a polynomial in $x_n$
with coefficients in $\A_n'$,
\begin{equation*}
  G=G_lx_n^l+G_{l-1}x_n^{l-1}+\cdots+G_1x_n+G_0,
\end{equation*}%
where $G_0,\dots,G_l\in\A_n'$. If $l=0$ then we are done. Let us suppose therefore that $l$ is minimal, but
$l>0$. We will show that this leads to a contradiction. Since each coefficient $G_i$ belongs to $\A_n'$, we have
that $G_i(t)=O(1)$.  Thus, the pole that $x_n(t)$ has, needs to be compensated by a zero in $G_l(t)$, i.e.,
$G_l(t)=O(t)$. We show that this implies that $G_lx_n\in\A_n'$. By Euclidean division in $\A_n'$ we can write $G_l$
as
\begin{equation}\label{sd_div}
  G_l=(1-x_{n+1}^2)K_1+(1-x_{n-1}^2)K_2+K_3,
\end{equation}
where $K_1,\,K_2$ and $K_3$ belong to $\A_n'$, and where $K_3$ is of degree $1$ at most in $x_{n+1}$ and $x_{n-1}$:
we can write $K_3$ as
\begin{equation*}
  K_3=\kappa_1(x_{n+1}+x_{n-1})+\kappa_2(1+x_{n+1}x_{n-1})+\kappa_3x_{n+1}+\kappa_4\\
\end{equation*}%
where $\kappa_1,\dots,\kappa_4$ are elements of $\A_n'$ that are independent of $x_{n+1}$ and $x_{n-1}$. Since
$G_l(t)=O(t)$ and $1-x_{n\pm1}^2(t)=O(t)$ it follows from (\ref{sd_div}) that $K_3(t)=O(t)$, and so that the
leading terms $\kappa_3^{(0)}$ and $\kappa_4^{(0)}$ of $\kappa_3(t)$ and $\kappa_4(t)$ satisfy
$\kappa_4^{(0)}=\e\kappa_3^{(0)}$. Since the leading terms $\e a_k$ of all $x_k(t)$, with
$k\in\Z\setminus\set{n-1,n,n+1}$, and the leading terms of $w_1(t)$ and $w_2(t)$ are all independent, even modulo
$\e$, it follows that $\kappa_4^{(0)}=\kappa_3^{(0)}=0$, as $\kappa_4$ and $\kappa_3$ are independent of
$x_{n\pm1}$. Using (\ref{wpmin}) it follows that
\begin{eqnarray*}
  G_lx_n&=&(1-x_{n+1}^2)x_nK_1+(1-x_{n-1}^2)x_nK_2+\kappa_1w_1+\kappa_2w_2\\
        &=&(w_2-x_{n+1}w_1)K_1+(w_2-x_{n-1}w_1)K_2+\kappa_1w_1+\kappa_2w_2,
\end{eqnarray*}
where $K_1,K_2,\kappa_1,\kappa_2\in\A_n'$, showing that $G_lx_n=G_l'\in\A_n'$, as promised. Then,
\begin{equation*}
  G=(G_l'+G_{l-1})x_n^{l-1}+\cdots+G_1x_n+G_0,
\end{equation*}%
with $G_l'+G_{l-1}\in\A_n'$. This contradicts the minimality of $l$. 
\qed
\end{proof}
\begin{lemma}\label{dep_self_lemma}
  For $k\neq n$, $\G_k(t):=\G_k(x(t);u(t))$ is of the form
  \begin{equation}
    \G_k(t)=\FF(a_{k-N},a_{k-N+1},\dots,a_{k+N},a_+,a_-)+O(t),
  \end{equation}
  i.e., the constant term in $\G_k(t)$ is a polynomial in the variables\footnote{Recall that $a_{n\pm1}=\mp1$ and
  that $a_n$ does not exist; so $a_{n\pm1}$ and $a_n$ may be thought of as being absent in the list. Thus, $a_\pm$
  is the natural substitute for $a_{n\pm1}$.}  $a_{k-N},$ $a_{k-N+1},$ $\dots,$ $a_{k+N},$ $a_+$ and $a_-$ only.
\end{lemma}
\begin{proof}
  According to (\ref{high_selfdual}), $\G_k$ depends on $x_{k-N},\dots,x_{k+N}$ only. For $k\neq n$ we know from
  Proposition \ref{self-dual_tan_prop} that $\G_k(t)=O(1)$, so that Proposition~\ref{pol_char_lemma} yields that
  $\G_k$ depends on $x_n$ through $w_1$ and $w_2$ only, i.e., $\G_k$ is a polynomial in $w_1,\,w_2$ and the $x_l$
  with $\vert k-l\vert\leq N$ and $l\neq n$.  Each of these variables is $O(1)$, so the constant term in $\G_k$ is
  a polynomial in their leading terms, which are the parameters $a_{k-N},$ $a_{k-N+1},\dots,$ $a_{k+N},$ $a_+$ and
  $a_-$ (see (\ref{ws})).
\qed
\end{proof}
It is clear that when $\vert k-n\vert>N$ then $\G_k(0)$ is independent of $a_+$ and $a_-$, as it cannot contain
$w_1$ or $w_2$. The following lemma deals with the case of $\G_n(t)$, which is slightly harder because $\G_n(t)$
develops a pole.
\begin{lemma}\label{GN_lemma}
  $\G_n(t):=\G_n(x(t);u(t))$ is of the form
\begin{equation*}
  \G_n(t)=\frac{\G_{n+1}^{(0)}-\G_{n-1}^{(0)}}{4t}+\FF(a_{n-N-1},\dots,a_{n+N+1},a_+,a_-) +O(t)
\end{equation*}%
  where $\FF$ is a polynomial in all its arguments, with $a_{n+N+1}$ and $a_{n-N-1}$ present (linearly).
\end{lemma}
\begin{proof}
  Consider the following alternative ways of writing $\G_n=\G_n(x;u)$,
\begin{equation}\label{trick}
  \G_n(x;u)=v_n H_n(x;u)+nx_n=x_nG_n(x;u)+H_n(x;u).
\end{equation}
$H_n$ is a polynomial in $x=(x_i)_{i\in\Z}$, because (\ref{G_eq}) implies that $H_n(x;u)=\V^u[x_n]$, and because
$\p x_n/\p t_i=\pb{x_n,H_i}$ is always divisible by $v_n$, see~(\ref{one_vec}). Also, we have put
$G_n(x;u):=n-x_nH_n(x;u)$ to obtain the second equality. The first equation in (\ref{trick}) implies that
$H_n(x(t);u(t))=O(t)$, since $\G_n(x(t);u(t))=O(t^{-1})$ and $x_n(t)=O(t^{-1})$, while
$v_n(t)=-1/(4t^2)+O(t^{-1})$. The second equation in (\ref{trick}) then allows us to conclude that
$G_n(x(t);u(t))=O(1)$, and hence also that $G_n(x(t);u)=O(1)$, since $u$ is an arbitrary vector of constants. Thus,
$G_n$ is, by Proposition \ref{pol_char_lemma}, an element of $\A_n'$, depending (linearly) on the parameters $u_i$.

\smallskip

Summarizing, the constant term in $\G_n(t)$ will be given by the constant term in $x_n(t)G_n(t)$, hence will depend
only on the first two terms $\e(1+(a_+-a_-)t)/(2t)$ of $x_n(t)$ and on the first two terms of $G_n(t)$, where
$G_n\in\A_n'$. The latter first two terms can depend only on the first two terms of the variables
$x_{n-N},\dots,x_{n+N},\,w_1$ and $w_2$ that appear in $G_n$; the first two terms of their series can be read off
from 
(\ref{ws}), yielding that the constant term in $\G_n(t)$ can only depend on
$a_{n-N-1},\dots,a_{n+N+1},a_+,a_-$. Notice that the only dependence on $a_{n-N-1}$ can come from the presence of
$x_{n-N}$, but (\ref{high_selfdual}) tells us that $x_{n-N}$ appears linearly in $\G_n$, and with a non-zero
coefficient. Therefore, the parameter $a_{n-N-1}$ is indeed present in the constant term in $\G_n$; similarly,
$a_{n+N+1}$ is also present. The leading term of $\G_n(t)$ was already determined in Proposition
\ref{self-dual_tan_prop}.
\qed
\end{proof}

\subsection{Parameter restriction}
We now show that we can tune the free parameters in the formal Laurent solution $x(t)$ of the self-dual Toeplitz lattice
in such a way that $\G_k(t)=0$ for all $k\in\Z$, as a formal series in $t$. As it turns out, it will be possible to
keep $2N-1$ parameters arbitrary, and the other ones are determined rationally in terms of these. Together with
time it means that the constructed solution depends on $2N$ free parameters, which is the maximum one can hope for
in an $2N+1$ step relation.
\begin{proposition}\label{sd_restr_prop}
  Keeping the $2N-1$ parameters $a_{n-2N},\dots,a_{n-2}$ arbitrary, the other parameters in the formal Laurent series
  $x(t)$, given by Proposition \ref{self_dual_laurent_prop}, can be chosen as rational functions of these
  parameters, so that $\G_k(t)=0$, as a formal series in $t$, for all $k\in\Z$.
\end{proposition}
\begin{table}[ht]
  \caption{Setting $\G_k(0)=0$ in the given order allows us to solve for all free parameters in the formal Laurent series,
  except for the $2N-1$ parameters $a_{n-2N},\dots,a_{n-2}$, that can be taken arbitrarily. We solve (linearly) for
  the underlined terms.\label{selfdual_table}}
  \begin{center}
  \renewcommand{\arraystretch}{1.5}
  \begin{tabular}{|c|c|c|c|c|}
    \hline 
    step &$\G_k$&$\G_k$ polynomial in& $\G_k^{(0)}$ polynomial in\\ 
    \hline 
      (1)&$\G_{n-N-1}$&$x_{n-2N-1},\dots,x_{n-1}$&$\underline{a_{n-2N-1}},\dots,{a_{n-1}}=1$\\ 
      (2)&$\G_{n-N-2}$&$x_{n-2N-2},\dots,x_{n-2}$&$\underline{a_{n-2N-2}},\dots,{a_{n-2}}$\\ 
      (3)&$\vdots$&$\vdots$&$\vdots$\\ \hline
      (4)&$\G_{n-N}$&$x_{n-2N},\dots,x_{n}$& ${a_{n-2N}},\dots,a_{n-2},\underline{a_-}$\\
      (5)&$\G_{n-N+1}$&$x_{n-2N+1},\dots,x_{n+1}$& ${a_{n-2N+1}},\dots,a_{n-2},a_-,\underline{a_+}$\\
      (6)&$\G_{n-N+2}$&$x_{n-2N+2},\dots,x_{n+2}$& ${a_{n-2N+2}},\dots,a_{n-2},a_\pm,\underline{a_{n+2}}$\\
      (7)&$\vdots$&$\vdots$&$\vdots$\\ 
      (8)&$\G_{n-1}$&$x_{n-N-1},\dots,x_{n+N-1}$&${a_{n-N-1}},\dots,a_{n-2},a_\pm,$\\ 
             &&&$a_{n+2},\dots,\underline{a_{n+N-1}}$\\
      (9)&$\G_{n+1}$&$x_{n-N+1},\dots,x_{n+N+1}$& ${a_{n-N+1}},\dots,a_{n-2},a_\pm$\\
             &&&$a_{n+2},\dots,\underline{a_{n+N}},\crossed{a_{n+N+1}}$\\ 
     (10)&$\G_{n}$&$x_{n-N},\dots,x_{n+N}$&${a_{n-N-1}},\dots,a_{n-2},a_\pm$\\ 
            &&&$a_{n+2},\dots,\underline{a_{n+N+1}}$\\
     (11)&$\G_{n+2}$&$x_{n-N+2},\dots,x_{n+N+2}$& ${a_{n-N+2}},\dots,a_{n-2},a_\pm$\\ 
            &&&$a_{n+2},\dots,\underline{a_{n+N+2}}$ \\ 
     (12)&$\vdots$&$\vdots$&$\vdots$\\ 
    \hline
  \end{tabular}
  \end{center}
\end{table}

\begin{proof}
In this proof we will assume that $N>1$. See Remark \ref{N=1rem} below for the adaption to the case
$N=1$. According to Proposition \ref{self-dual_tan_prop}, it suffices to determine the parameters in the series
$x(t)$ so that $\G_k^{(0)}$, the constant term in $\G_k(t)$, is zero, for all $k\in\Z$. Thus, we need to write
$\G_k^{(0)}$ in terms of the parameters in the series $x(t)$. We do this for the different values of $k$ in a very
specific order, as indicated in Table 1. The second column indicates which $\G_k$ we consider; it is easy to see
that we consider all of them (exactly once); it is understood that steps (6)--(8) are absent when $N=2$.  We know
from (\ref{high_selfdual}) that for any $k\in\Z$, $\G_k$ depends only on the variables
$x_{k-N},\,x_{k-N+1},\dots,x_{k+N}$, which yields the third column. It is important to point out that the two
written variables, which are the extremal terms, are actually present in $\G_k$, and that these two variables
appear linearly (see Proposition \ref{app_prop} in the Appendix).

\smallskip

The delicate step is in obtaining the last column; the information displayed in it contains the
parameters\footnote{Besides the constants $u_1,\dots,u_N$ that define $P$.}  that may appear in $\G_k^{(0)}$, where
the underlined term actually does appear, and it appears linearly. Before validating this column in each of the
steps, let us first point out how the proposition follows from it. Precisely, we can in each step solve for one of
the underlined parameters in terms of the nonunderlined parameters, as the underlined parameter appears linearly in
the equation $\G_k^{(0)}=0$. Using the previous steps, this yields (using the previous steps) inductively a
rational formula for each of the parameters, in terms of $a_{n-2N},\dots,a_{n-2}$, which remain free. In fact, the
variables $a_{n-2N-i}$, with $i>0$ are determined in steps (1) -- (3); $a_{n-1}=-a_{n+1}=1$ while $a_n$ does not
exist; the variables $a_{n+i+1}$ with $i>0$ are determined in steps (6) -- (12); the only other variables are $a_-$
and $a_+$, which are determined in steps (4) and (5).

\smallskip

We now show that in each step the parameters that are indiciated in the fourth column of the table appear indeed
(linearly) in $\G_k^{(0)}$. This is done by carefully using the leading terms of $\G_k$, as given by
Proposition~\ref{app_prop}. As a general remark, notice that~(\ref{high_selfdual}) implies that $\G_k$ contains the
variables $x_{k-N}$ and $x_{k+N}$ linearly, but that the behaviour of its coefficients $\prod_{i=0}^{N-1}v_{k+i}$
and $\prod_{i=0}^{N-1}v_{k-i}$, evaluated at $t$, depends on $k$, as given in (\ref{v_k_sd}).

\smallskip

For step (1) we have that $x_{n-2N-1}(t), $ $\dots, x_{n-1}(t)$ have no pole in $t$, so that only their leading
coefficients, the parameters ${a_{n-2N-1}},\dots,a_{n-2},$ ${a_{n-1}}=1$, can appear. Since $x_{n-2N-1}$ appears
(linearly) in $\G_{n-N-1}$, with a coefficient $u_N\prod_{i=1}^{N}v_{n-N-i}$ that is non-vanishing for $t=0$,
namely $\prod_{i=1}^{N}v_{n-N-i}(0)= \prod_{i=1}^{N}\check a_{n-N-i}$, the parameter $a_{n-2N-1}$ appears
(linearly) in $\G_{n-N-1}^{(0)}$.  The same argument works in steps (2) and (3). Step (4) is more interesting
because it involves $x_n$ (linearly). However, $x_n$ appears only in the leading term of $\G_{n-N}$, which we can
write, using $w_-=x_nv_{n-1}$, as
\begin{equation}\label{sd_step4}
  u_Nx_n\prod_{i=0}^{N-1}v_{n-N+i}=u_Nw_-\prod_{i=0}^{N-2}v_{n-N+i},\qquad u_N\neq0.
\end{equation}
Now $w_-(t)=4\e a_-+O(t)$, and the other factors in (\ref{sd_step4}) are finite, non-vanishing, which yields the
proposed dependence on the parameters in step~(4). For step (5), $x_n$ may be present in other terms than the
leading term in $\G_{n-N+1}$, but in view of Proposition \ref{pol_char_lemma}, $\G_{n-N+1}\in\A_n'$ is a polynomial
in $x_{n-2N+1},\dots,x_{n-1},x_{n+1}$ and in $w_1$ and $w_2$ only. Since their series do not have a pole for $t=0$,
we get an eventual dependence on $a_+$ and $a_-$, besides the parameters $a_{n-2N+1},\dots,a_{n-2}$. Let us show
that $a_+$ actually appears. The leading term in $\G_{n-N+1}$ is, according to (\ref{high_selfdual}),
\begin{equation*}
  u_Nx_{n+1}v_nv_{n-1}\prod_{i=n-N+1}^{n-2}v_i.
\end{equation*}%
Since it is the only term in $\G_{n-N+1}$ that contains $x_{n+1}$ we can write $\G_{n-N+1}=P_1+P_2$, where
\begin{equation*}
  P_1=u_N(x_{n+1}+x_{n-1})v_nv_{n-1}\prod_{i=n-N+1}^{n-2}v_i,
\end{equation*}%
and $P_2$ is independent of $x_{n+1}$, so $P_2$ depends only on $x_{n-2N+1},\dots,x_n$. Now $P_1(t)=O(1)$, since
\begin{equation*}
   x_{n+1}(t)+x_{n-1}(t)=O(t),\qquad v_n(t)=O(t^{-2}),\qquad v_{n-1}(t)=O(t),
\end{equation*}%
while the other $v_i(t)$ that appear in $P_1(t)$ are $O(1)$.  Since $\G_{n-N+1}(t)=O(1)$ this implies that
$P_2(t)=O(1)$, so that $P_2$ satisfies the hypothesis of Proposition \ref{pol_char_lemma}; since $P_2$ is
independent of $x_{n+1}$ we may conclude, as in step (4), that $P_2$ is independent of $a_+$. On the other hand
$P_1(0)$ depends (linearly) on $a_+$, as
\begin{equation*}
  (x_{n+1}(t)+x_{n-1}(t))v_n(t)v_{n-1}(t)=8\e a_-(a_-+a_+)+O(t).
\end{equation*}%
The conclusion is that $\G_{n-N+1}^{(0)}=P_1(0)+P_2(0)$ depends (linearly) on $a_+$.

\smallskip

We are at step (6). Skip this step and steps (7) and (8) when $N=2$.  Proposition \ref{pol_char_lemma} implies that
$\G_{n-N+2}^{(0)}$ can only depend on the proposed parameters, and that the dependence comes from the constant
terms of the series in (\ref{ws}). The dependence of $\G_{n-N+2}^{(0)}$ on $a_{n+2}$ comes only from the leading
term $u_Nx_{n+2}v_{n+1}v_nv_{n-1}\prod_{i=0}^{N-4} v_{n-N+2+i}$ which, at $t$, is $O(1)$, since
the product $v_{n+1}(t)v_n(t)v_{n-1}(t)=O(1)$ and non-vanishing. It follows that $\G_{n-N+2}^{(0)}$ depends on $a_{n+2}$
(linearly). The same happens in steps (7) and (8), as the leading term will always contain the product
$v_{n+1}v_nv_{n-1}$ which is finite and non-zero for $t=0$.

\smallskip

A new phenomenon arises in step (9). Notice that we have moved to $\G_{n+1}$, keeping $\G_n$ for step (10). The
leading term of $\G_{n+1}$ is
\begin{equation*}
  u_Nx_{n+N+1}\prod_{i=1}^{N}v_{n+i},
\end{equation*}%
which does not contribute to $\G_{n+1}^{(0)}$, since $v_{n+1}(t)=O(t)$, while all other factors in this term are
finite in $t$. Therefore, $\G_{n+1}^{(0)}$ is independent of $a_{n+N+1}$. To show that $\G_{n+1}^{(0)}$ depends on
$a_{n+N}$ we need to investigate the next term in $\G_{n+1}$, the one that contains $x_{n+N}$, because it is the
only one that might lead to a dependence on $a_{n+N}$. According to Proposition \ref{app_prop}, this term consists of
the following three pieces,
\begin{eqn}{three_pieces}
  &\displaystyle u_{N-1}x_{n+N}\prod_{i=0}^{N-2}v_{n+1+i}-u_Nx^2_{n+N}x_{n+N-1} \prod_{i=0}^{N-2}v_{n+1+i}\\
  &\displaystyle -2u_Nx_{n+N}\prod_{i=0}^{N-2}v_{n+1+i}\sum_{j=0}^{N-2}x_{n+j+1}x_{n+j}.
\end{eqn}%
The two terms on the first line of (\ref{three_pieces}) do not contribute to $\G_{n+1}^{(0)}$, again because both
terms contain $v_{n+1}$, and all other terms are finite for $t=0$. The third term however does contribute, when
$j=0$, as $x_n(t)v_{n+1}(t)\sim a_++O(t)$; moreover, this term is the only one that involves $a_{n+N}$, so that the
latter parameter appears (linearly) in $\G_{n+1}^{(0)}$. For step (10) the presence of $a_{n+N+1}$ was established
in Lemma~\ref{GN_lemma}. Starting from step (11) the leading coefficients do not contain $v_{n\pm1}$ or $v_n$
anymore, so that everything goes smoothly.
\qed
\end{proof}
%
%
\begin{remark}\label{N=1rem}
When $N=1$ the polynomial that defines the recursion relation reduces to 
\begin{equation*}
  \G_k=kx_k+u_1(1-x_k^2)(x_{k+1}+x_{k-1}).
\end{equation*}%
Steps (4)--(9) then get replaced by two steps in which we consider $\G_{n\pm1}$, which allows us to determine
$a_\pm$. Indeed, substituting the series $x(t)$ in $\G_{n\pm1}$ yields for the leading term ($t=0$):
\begin{equation*}
  (n\pm1)+4u_1a_\pm=0.  
\end{equation*}%
The other parameters are determined as in the general case.
\end{remark}
%
\section{Restricting the formal Laurent solutions: the general case}
\label{conf_general_sec}
In this section we will do a similar analysis as the one that has been done for the case of the self-dual Toeplitz
lattice in Section \ref{conf_self-dual_sec}.
\subsection{Structure of the polynomials $\G_k$ and $\Gt_k$}
We first investigate on which parameters the leading term(s) in the polynomials $\G_k$ and $\Gt_k$ depends on the
free parameters. We denote by $\A$ the algebra of all polynomials in the variables $x_i$ and $y_i$, where $i\in\Z$,
while $\A_n$ stands for the subalgebra of $\A$ that consists of all polynomials that do not depend on $x_n$ and on
$y_n$. Consider the following four polynomials\footnote{Recall that $v_i:=1-x_iy_i$ in the case of the general
Toeplitz lattice, and that $\sigma$ denotes the involution that permutes all $x_i\leftrightarrow y_i$.}
\begin{equation}\label{w_gen_def}
  \begin{array}{rclrcl}
    w_1&=&x_ny_{n-1}+y_nx_{n+1},\qquad&w_2&=&x_n+x_{n-1}y_nx_{n+1},\\  
    w_1^\s&=&y_nx_{n-1}+x_ny_{n+1},\qquad&w_2^\s&=&y_n+y_{n-1}x_ny_{n+1}.
  \end{array}
\end{equation}
For future use, observe that these polynomials are linked by the following identity:
\begin{equation}\label{my_id}
  x_n(w_2^\s-y_{n-1}w_1^\s)=  y_n(w_2-x_{n-1}w_1),
\end{equation}
in fact both expressions in (\ref{my_id}) are equal to $x_ny_nv_{n-1}$. We denote by $\A_n'$ the subalgebra of $\A$
that consists of all polynomials that can be written in terms of these four polynomials, besides all $x_i$ and
$y_i$, with $i\neq n$. The polynomials $w$ have the following series in $t$, when the first few\footnote{A priori,
one needs to compute an extra term in the series $z_k(t)$ (see Proposition \ref{general_laurent_prop}) in order to
find the shown terms in (\ref{w_gen_lau}). After Proposition \ref{gen_pol_char_lemma} we will however show how such
a cumbersome can be avoided.} terms of the series $x_i(t)$ and $y_i(t)$ that are constructed in Proposition
\ref{general_laurent_prop}, are substituted in them.
\begin{eqn}{w_gen_lau}
    w_1(t)&=&{\Omega}{b_{n-1}}-a_-+(a_+a_{n+2}b_{n-1}-a_-a_{n-1}b_{n-2})t+O(t^2),\\
    w_2(t)&=&\Omega+(a_+a_{n+2}+a_-a_{n-2})t+O(t^2),
\end{eqn}%
where
\begin{equation*}
  \Omega:=\frac{a_{n-1}a_{n+1}}{(a_{n+1}-a_{n-1})^2} \left(a_{n-1}(2a-a_+)-a_{n+1}(2a-a_-)\right).
\end{equation*}%
The formal Laurent series for the other polynomials in (\ref{w_gen_def}) is found from it by using the automorphism
$\s$ (see (\ref{s_ext})), which yields in particular
\begin{equation}\label{somega}
  \sigma(\Omega)={\Omega}{b_{n-1}b_{n+1}}+{a_+}{b_{n-1}} -{a_-}{b_{n+1}}.
\end{equation}
It follows that if $G\in\A_n'$ then $G(t)=O(1)$, where $G(t):=G(x(t),y(t))$, with $x(t)$ and $y(t)$ as above. We
will show that the converse is also true, so that the algebra $\A_n'$ plays in the general case a similar r\^ole as
in the self-dual case. For this we need the following lemma.
\begin{lemma}\label{my_lemma}
  Let $G$ be a polynomial in $\A_n'$ that is independent of $w_2$ and none of whose terms contains $x_{n+1}y_{n+1}$
  or $x_{n-1}y_{n-1}$. If $G(t)=O(t)$ then $G=0$, as a formal series in $t$.
\end{lemma}
\begin{proof}
It follows from (\ref{w_gen_lau}) that
\begin{equation*}\renewcommand{\arraystretch}{1.5}
  \begin{matrix}{c}
    w_1(0)\\w_1^\s(0)\\w_2^\s(0)
  \end{matrix}
  =\frac T{(a_{n+1}-a_{n-1})^2}
  \begin{matrix}{c}
    a(a_{n-1}-a_{n+1})\\ a_+a_{n+1}\\ a_-a_{n-1}
  \end{matrix}
\end{equation*}%
where
\begin{equation*}
  \renewcommand{\arraystretch}{1.5}
  T:=\begin{matrix}{ccc}
    2a_{n+1}&-a_{n-1}&2a_{n+1}-a_{n-1}\\
    2a_{n-1}&a_{n+1}-2a_{n-1}&a_{n+1}\\
    2&\frac{1}{a_{n-1}}(a_{n+1}-2a_{n-1})&\frac{1}{a_{n+1}}(2a_{n+1}-a_{n-1})
  \end{matrix}.
\end{equation*}%
$T$ is an invertible matrix, since $\det T=-2(a_{n-1}-a_{n+1})^4/(a_{n-1}a_{n+1})$. Let $G$ be a polynomial in
$\A_n'$ that is independent of $w_2$ and suppose that $G(0)=0$. We write
$G=\sum_{ijk}g_{ijk}w_1^i(w_1^\s)^j(w_2^\s)^k$, where $g_{ijk}$ is a polynomial in the variables $x_k$ and $y_k$
with $k\neq n$ only. Notice that $g_{ijk}(0)$ is independent of $a,\,a_+$ and $a_-$. Therefore, the fact that $T$
is invertible and that $a,\,a_+$ and $a_-$ are independent free variables implies that $g_{ijk}(t)=O(t)$ for any
$i,j,k$. If we assume now in addition that $g_{ijk}$ does not contain either product $x_{n+1}y_{n+1}$ or
$x_{n-1}y_{n-1}$ then it is clear that $g_{ijk}=0$ since the leading terms $a_k$ of $x_k$ and $b_k$ of $y_k$ are
independent ($k\neq n$), except that $a_{n+1}b_{n+1}=1=a_{n-1}b_{n-1}$.
\qed
\end{proof}
\begin{proposition}\label{gen_pol_char_lemma}
  For $G\in\A$, let $G(t):=G(x(t),y(t))$, where $(x(t),y(t))$ is the formal Laurent solution to the first vector field of
  the Toeplitz lattice, constructed in Proposition \ref{general_laurent_prop}. If $G(t)=O(1)$ then $G\in\A_n'$,
  i.e., $G$ depends only on $x_n$ and $y_n$ through the polynomials $w_1,\,w_2,\,w_1^\s$ and $w_2^\s.$
\end{proposition}
\begin{proof}
Given $G\in\A$ we may write $G$ as a polynomial in $x_n$ and $y_n$, with coefficients in $\A_n'$; in fact, writing
$x_n=w_2-x_{n-1}y_nx_{n+1}$ we may assume that $G$ is independent of $x_n$ and we write
\begin{equation*}
  G=G_ly_n^l+G_{l-1}y_n^{l-1}+\cdots+G_1y_n+G_0,
\end{equation*}%
where $G_0,\dots,G_l\in\A_n'$. We suppose that this is done in such a way that $l$ is minimal. If $l=0$ then
$G\in\A_n'$ and we are done; assume therefore that $l>1$.  We will show that $G_ly_n\in\A_n'$, which is in
contradiction with the minimality of $l$, like in the self-dual case. We first show that we may assume that $w_2$
is absent in $G_ly_n$. If we substitute $x_n=w_2-x_{n-1}y_nx_{n+1}$ in the identity (\ref{my_id}) then we find
\begin{equation*}
  y_nw_2=w_2(w_2^\s-y_{n-1}w_1^\s)+y_n(w_1x_{n-1}+x_{n-1}x_{n+1}(y_{n-1}w_1^\s-w_2^\s)),
\end{equation*}%
which allows us to replace any term in $G_ly_n$ that contains $w_2$, or a power of it, by a term of lower degree in
$w_2$, at the cost of changing $G_{l-1}$, so that we can eventually remove $w_2$ entirely from the leading
coefficient $G_l$. Assuming that $G_l$ does not depend on $w_2$ we perform an Euclidean division in $\A_n'$,
\begin{equation}\label{glexp}
  G_l=(1-x_{n-1}y_{n-1})K_1+(1-x_{n+1}y_{n+1})K_2+K_3,
\end{equation}
where $K_1,K_2$ and $K_3$ belong to $\A_n'$, with $K_3$ independent of $w_2$ and not containing $x_{n-1}y_{n-1}$ or
$x_{n+1}y_{n+1}$.

Assume now that $G(t)=O(1)$. Since all $G_i(t)$ are $O(1)$, as $G_i\in\A_n'$, we must have that $G_l(t)=O(t)$, as
$y_n(t)$ has a pole. Then (\ref{glexp}) implies that $K_3(t)=O(t)$, since
$1-x_{n\pm1}(t)y_{n\pm1}(t)=v_{n\pm1}(t)=O(t)$. This means that $K_3$ satisfies the conditions of Lemma
\ref{my_lemma}, hence that $K_3=0$. The identities
\begin{eqnarray*}
  (1-x_{n-1}y_{n-1})y_n=w_2^\s-y_{n-1}w_1^\s\in\A_n'\\
  (1-x_{n+1}y_{n+1})y_n=w_2^\s-y_{n+1}w_1\in\A_n'
\end{eqnarray*}
then imply that $G_ly_n\in\A_n'$, as was to be shown.
\qed
\end{proof}
As a first application of this proposition, we show how the shown terms in (\ref{w_gen_lau}) can easily be
computed. Since $w_i(t)=O(1)$ we also have $\dot w_i(t)=O(1)$ for $i=1,2$. By Proposition \ref{gen_pol_char_lemma},
$\dot w_i\in \A_n'$, in fact
\begin{eqnarray*}
  \dot w_1&=&(x_ny_{n-1}+y_nx_{n+1})^\cdot\\
           &=&y_{n-2}x_{n-1}x_{n+1}w_-^\s+x_{n+2}w_+^\s-y_{n-2}v_{n-1}w_2+v_{n-1}-v_{n+1},\\
  \dot w_2&=&\dot x_n+(x_{n-1}y_nx_{n+1})^\cdot\\
           &=&x_{n+2}x_{n-1}w^\s_+-x_{n-2}x_{n+1}w_-^\s+x_{n+1}v_{n-1}-x_{n-1}v_{n+1},
\end{eqnarray*}%
where $w_\pm^\s:=v_{n\pm1}y_n$, with $w_\pm^\s(0)=\pm a_\pm b_{n\mp1}+O(t)$. Since $v_{n\pm1}(0)=0$, it follows that
\begin{eqnarray*}
  \dot w_1(0)&=&b_{n-2}a_{n-1}a_{n+1}w_-^\s(0)+a_{n+2}w_+^\s(0)=a_+a_{n+2}b_{n-1}-a_-a_{n-1}b_{n-2},\\
  \dot w_2(0)&=&a_{n+2}a_{n-1}w^\s_+(0)-a_{n-2}a_{n+1}w_-^\s(0)=a_{n+2}a_++a_{n-2}a_-,
\end{eqnarray*}
which yield after integration the linear terms in (\ref{w_gen_lau}). The same formulas can be used to show that
$w_1^{(2)}$ and $w_2^{(2)}$, which are the $t^2$ terms in $w_1(t)$ and in $w_2(t)$, depend only on the parameters
$c_{n-3},\dots,c_{n+3},\,a_+,\,a_-$ and $a$; the precise formula will not be needed, except that they depend on
$c_{n+3}$ as follows:
\begin{eqn}{wi_spec}
  w_1^{(2)}&=&x_{n+2}^{(1)}w_+^\s(0)/2+\cdots=a_{n+3}a_+b_{n-1}\check c_{n+2}/2+\cdots,\\
  w_2^{(2)}&=&x_{n+2}^{(1)}x_{n-1}(0)w_+^\s(0)/2+\cdots=a_{n+3}a_+\check c_{n+2}/2+\cdots,
\end{eqn}
where the dots are independent of $a_{n+3}$ (and of $b_{n+3}$).
%

\smallskip

The following lemma is the analog of Lemma \ref{dep_self_lemma} and is proven in exactly the same way.
\begin{lemma}\label{dep_gen_lemma}
  If $k\neq n$, then the series $\G_k(t):=\G_k(x(t),y(t);u(t))$ and
  $\Gt_k(t):=\Gt_k(x(t),y(t);u(t))$ are of the form
  \begin{eqn*}
    \G_k(t)=\FF(a_{k-N},c_{k-N+1},\dots,c_{k+N-1},a_{k+N},a_\pm,a)+O(t),\\
    \Gt_k(t)=\tilde\FF(b_{k-N},c_{k-N+1},\dots,c_{k+N-1},b_{k+N},a_\pm,a)+O(t),
  \end{eqn*}%
  where we recall that $c_i=(a_i,b_i)$ and that $a_{n\pm1}b_{n\pm1}=1$, and  $\FF,\tilde\FF$ are polynomials in their
  arguments. 
\end{lemma}
%
For $k=n$ the corresponding result is more complicated and the method of proof is different from the one in the
self-dual case (Lemma \ref{GN_lemma}).
\begin{lemma}\label{step10lemma}
  The constant terms $\G_n^{(0)}$ and $\Gt_n^{(0)}$ are of the form
\begin{equation*}
  \left(
  \begin{array}{cc}
    \G_n^{(0)}\\ \Gt_n^{(0)}
  \end{array}
  \right)
  =A\left(
  \begin{array}{cc}
    a_{n+N+1}\\ b_{n+N+1}
  \end{array}
  \right)
  +\FF(c_{n-N-1},\dots,c_{n+N},a_\pm,a),
\end{equation*}%
  where $A$ is an invertible $2\times 2$ matrix and $\FF$ is a polynomial 2-vector that depends on the listed free
  parameters only. See Proposition \ref{tang_prop} for the leading terms of $\G_n(t)$ and $\Gt_n(t)$.
\end{lemma}
\begin{proof}
We will assume in our proof that $N>2$, see Remark \ref{rem_n=2} below. The proof is based on the explicit
expression for $\G_n$ that is given in Proposition \ref{app_prop} (see the Appendix), which we write in the form
$\G_n=v_nH_n+nx_n$, where
\begin{eqn*}
  H_n&=&u_Nx_{n+N}\prod_{i=1}^{N-1}v_{n+i}-u_Nx_{n+N-1}^2y_{n+N-2} \prod_{i=1}^{N-2}v_{n+i}\\
     &&\,-u_Nx_{n+N-1}\left(x_ny_{n-1}+2\sum_{j=1}^{N-2}x_{n+j}y_{n+j-1}\right)\prod_{i=1}^{N-2}v_{n+i}\\ 
     &&+(u_{N-1}x_{n+N-1}-u_{-N}y_{n+N-1}x_{n-1}x_n)\prod_{i=1}^{N-2}v_{n+i}\\
     &&+\,\FF(x_{n-N+1},\dots,x_{n+N-2},y_{n-N+2},\dots,y_{n+N-2})\\
     &&-\,(u_Nx_nx_{n+1}y_{n-N+1}-u_{-N}x_{n-N}v_{n-N+1})\prod_{i=1}^{N-2}v_{n-i}.
\end{eqn*}%
Our first claim is that $\FF\in\A_n'$. Since $\G_n(t)$ and $v_n(t)$ have a double pole, while $x_n(t)$ has a simple
pole, $H_n(t)=O(1)$. The terms in the above expression that do not involve $x_n$ or $y_n$ are also $O(1)$, because
$x_k(t)=O(1)$ and $y_k(t)=O(1)$ for $k\neq n$. There are a few terms that contain $x_n$ or $y_n$ (linearly), but
they are all of the form $x_nv_{n+1}$, $y_nv_{n+1}$ or $x_nv_{n-1}$, which are both $O(1)$. It follows that
$\FF(t;u(t))=O(1)$, and hence that $\FF(t;u)=O(1)$. Thinking of $u$ as constants we have, in view of Proposition
\ref{gen_pol_char_lemma}, that $\FF\in\A_n'$.

Since $v_n(t)$ has a double pole, only the first three terms of $v_n(t)$ and of $\FF(t)$ can contribute to the
constant term in $v_n(t)\FF(t)$; in view of Table~\ref{dependence_table}, this contribution can only yield
a dependence on the parameters $c_{n-N-1},\dots,c_{n+N},a_\pm$ and~$a$.

\smallskip

We now turn to the other terms in $H_n$ and we use their explicit form to show that they only depend on the listed
parameters. Let us first consider the following terms that do not involve $x_n$ or $y_n$,
\begin{eqn}{fterms}
  &&-\left(u_Nx_{n+N-1}^2y_{n+N-2}+2u_Nx_{n+N-1}\sum_{j=2}^{N-2}x_{n+j}y_{n+j-1}\right.\\
  &&\qquad\qquad-u_{N-1}x_{n+N-1}\Big)\prod_{i=1}^{N-2}v_{n+i}+\,u_{-N}x_{n-N}\prod_{i=1}^{N-1}v_{n-i}.
\end{eqn}%
Since $v_{n\pm i}$ has a simple zero for $i=1$ and is $O(1)$ for $i>1$ we have that $\prod_{i=1}^{N-2}v_{n+i}$ and
$\prod_{i=1}^{N-1}v_{n-i}$ have a simple zero, so we only need to look for the parameters that appear in the first
two terms of the coefficients. The former add nothing new to the above parameter list. For the coefficients of the
first one for example, we read off from Table \ref{dependence_table} that the constant and linear terms of
$x_{n+N-1}^2(t)y_{n+N-2}(t)$ only depend on $a_{n+N},c_{n+N-1},c_{n+N-2}$ and $b_{n+N-3}$, which falls inside the
proposed limits. Notice in particular that neither $a_{n+N+1}$ nor $b_{n+N+1}$ appear in this term. We arrive
similarly at the same conclusion for the other three terms in (\ref{fterms}). Notice that the lowest free parameter
that appears is $a_{n-N-1}$; it comes from the last term in (\ref{fterms}).

\smallskip

We now get to the terms that contain $x_n$ or $y_n$. As we already noticed these terms always come with $v_{n+1}$
or $v_{n-1}$. As $x_n(t)v_{n\pm1}(t)=O(1)$ we must investigate the first three terms in the remaining factors. For
the term
\begin{equation*}
  -u_Nx_nv_{n-1}x_{n+1}y_{n-N+1}\prod_{i=2}^{N-2}v_{n-i}
\end{equation*}%
we need to look at $x_{n+1}y_{n-N+1}\prod_{i=2}^{N-2}v_{n-i}$, which yields terms with a low index, the lowest
coming from the coefficient in $t^2$ in $y_{n-N+1}(t)$, to wit $b_{n-N-1}$ and $a_{n-N}$. The other three terms
that involve $x_n$ or $y_n$ can be written as
\begin{equation*}
  B:=-
  \renewcommand{\arraystretch}{1.5}
  \begin{matrix}{c}
     x_nv_{n+1}\left(u_Nx_{n+N-1}y_{n-1}+u_{-N}y_{n+N-1}x_{n-1}\right)\\
     +2u_Ny_nv_{n+1}x_{n+N-1}x_{n+1}
  \end{matrix}
  \prod_{i=2}^{N-2}v_{n+i}.
\end{equation*}%
Again, since $v_n$ has a double pole the first three terms in $B(t)=B_à+B_1t+B_2t^2+O(t^3)$ will contribute to the
constant term in $v_n(t)B(t)$. It is clear that $B_2$ will contain $a_{n+N+1}$, coming from $x_{n+N-1}^{(2)}$ and
$b_{n+N+1}$, coming from $y_{n+N-1}$. To know the precise value, it suffices to substitute the relevant
coefficients of the formal Laurent series $x(t),y(t)$ in the following part of $B_2$,
\begin{equation*}
  -
  \begin{matrix}{c}
    (x_nv_{n+1})^{(0)}\left(u_Nx^{(2)}_{n+N-1}y^{(0)}_{n-1}+u_{-N}y^{(2)}_{n+N-1}x^{(0)}_{n-1}\right)\\
    +2u_N(y_nv_{n+1})^{(0)}x_{n+N-1}^{(2)}x_{n+1}^{(0)}    
  \end{matrix}
  \prod_{i=2}^{N-2}v_{n+i}^{(0)},
\end{equation*}%
which gives, by using Proposition \ref{general_laurent_prop}, and in particular $-(x_nv_{n+1})^{(0)}=a_+a_{n+1}$
and $-(y_nv_{n+1})^{(0)}=-a_+b_{n-1}$,
\begin{equation}\label{partA}
  -\frac{a_+a_{n+1}}2(u_Na_{n+N+1}b_{n-1}-u_{-N}a_{n-1}b_{n+N+1})\prod_{i=2}^{N} \check c_{n+i} +\cdots,
\end{equation}
where the dots are independent of $a_{n+N+1}$ and $b_{n+N+1}$. There remains one term in $H_n$, namely the leading
term $C:=u_Nx_{n+N}\prod_{i=1}^{N-1}v_{n+i}$. It does not involve $x_n$ but does involve $v_{n+1}$, which will also
lead to a dependence on $a_{n+N+1}$. Writing $C(t)=C_1t+C_2t^2+O(t^3)$ we have that
\begin{eqn*}
  C_2&=&u_Nx_{n+N}^{(1)}v_{n+1}^{(1)}\prod_{i=2}^{N-1}v_{n+i}^{(0)}\\
     &=&u_Na_{n+N+1}a_+(a_{n+1}-a_{n-1})b_{n-1}\prod_{i=2}^{N}\check c_{n+i}+\cdots,
\end{eqn*}%
where the dots are again independent of $a_{n+N+1}$ and $b_{n+N+1}$. Summing up, we have that the leading terms in
$\G_n^{(0)}$ are given by
\begin{equation*}
  \frac{a_+v_n^{(0)}}2\left(u_N(a_{n+1}-2a_{n-1})b_{n-1}a_{n+N+1}+u_{-N}a_{n+1}a_{n-1}b_{n+N+1}\right)
  \prod_{i=2}^{N}\check c_{n+i}.
\end{equation*}%
By duality, the leading terms in $\Gt_n^{(0)}$ are given by
\begin{equation*}
  -\frac{a_+v_n^{(0)}}{2a_{n-1}}\left(u_{-N}(a_{n-1}-2a_{n+1})b_{n+N+1}+{u_N}b_{n-1}a_{n+N+1}\right)
  \prod_{i=2}^{N}\check c_{n+i}.
\end{equation*}%
We may conclude that 
\begin{equation}\label{rec_gen}
  \left(
  \begin{array}{cc}
    \G_n^{(0)}\\ \Gt_n^{(0)}
  \end{array}
  \right)
  =A\left(
  \begin{array}{cc}
    a_{n+N+1}\\ b_{n+N+1}
  \end{array}
  \right)
  +\FF(c_{n-N-1},\dots,c_{n+N},a_\pm,a),
\end{equation}%
where 
\begin{equation*}
  A=\frac{a_+a_{n+1}}{2(a_{n-1}-a_{n+1})^2}  
  \renewcommand{\arraystretch}{1.7}
  \begin{matrix}{cc}
    (a_{n+1}-2a_{n-1})u_N\ &a_{n+1}a_{n-1}^2u_{-N}\\
    \displaystyle-\frac{u_N}{a_{n-1}}&(2a_{n+1}-a_{n-1})u_{-N}
  \end{matrix}
  \prod_{i=2}^{N}\check c_{n+i}.
\end{equation*}%
Since 
\begin{equation*}
  \det A=\frac{u_Nu_{-N}}2\left(\frac{a_+a_{n+1}}{a_{n+1}-a_{n-1}}\prod_{i=2}^{N}\check c_{n+i}\right)^2,
\end{equation*}%
$A$ is invertible.  
\qed
\end{proof}
\begin{remark}\label{rem_n=2}
The above proof breaks down at several places when $N=2$. The polynomial $H_n$ then reduces to
\begin{eqn}{HnN=2}
  H_n&=&u_2(x_{n+2}v_{n+1}-x_{n+1}w_1)+u_1x_{n+1}\\
     &&+\,u_{-2}(x_{n-2}v_{n-1}-x_{n-1}w_1^\s)+u_{-1}x_{n-1}.
\end{eqn}%
Using (\ref{wi_spec}) and Proposition \ref{general_laurent_prop} we find that $H_n$ depends in the following way on
$a_{n+3}$ and $b_{n+3}$,
\begin{eqnarray*}
  \lefteqn{u_2(x_{n+2}^{(1)}v_{n+1}^{(1)}-x_{n+1}^{(0)}w_1^{(2)})-u_{-2}x_{n-1}^{(0)}w_1^{\s(2)}}\hskip2cm\\
  &=&\frac{a_+\check c_{n+2}}{2a_{n-1}}(u_2(a_{n+1}-2a_{n-1})a_{n+3}+u_{-2}a_{n+1}a_{n-1}^2b_{n+3}).
\end{eqnarray*}
It leads as in the case $N>2$ to (\ref{rec_gen}), with precisely the same matrix $A$.
\end{remark}
\subsection{Parameter restriction}

The parameter restriction works more or less like in the self-dual case, the main difference coming from the fact
that in the self-dual case we had to put all $\G_k^{(0)}=0$, while in the general case the tangency condition is
equivalent to
\begin{enumerate}
  \item $\G_k(t)=O(t)$ and $\Gt_k(t)=O(t)$ for all $k$ with $k\neq n+1$;
  \item $\G_{n-1}(t)=O(t^2)$;
  \item $\G_{n+1}(t)=O(t)$.
\end{enumerate}
In a sense, the condition $\G_{n-1}(t)=O(t^2)$ replaces the condition $\Gt_{n+1}(t)=O(t)$, which is redundant
because it is a consequence of the other conditions (see Proposition \ref{tang_gen_prop}).
\begin{proposition}
  Keeping the $4N-1$ parameters\footnote{Recall that $c_k=(a_k,b_k)$ and that $a_{n\pm1}b_{n\pm1}=1$.}
  $c_{n-2N},\dots,c_{n-2},a_{n-1}$ arbitrary, the other parameters in the formal Laurent series $(x(t),y(t))$, given by
  Proposition \ref{general_laurent_prop}, can be chosen as rational functions of these parameters, so that
  $\G_k(t)=0$ and $\Gt_k(t)=0$, identically in $t$, for all $k\in\Z$.
\end{proposition}
\begin{table}[ht]
  \caption{The tangency condition allows us to solve for all free parameters in the formal Laurent series, except for the
  $4N-1$ parameters $c_{n-2N},\dots,c_{n-2},a_{n-1}$, that can be taken arbitrarily. The equations can be solved
  linearly for the underlined terms. \label{general_table}}
  \begin{center}
  \renewcommand{\arraystretch}{1.5}
  \begin{tabular}{|c|c|c|c|c|}
    \hline 
   step &$\Delta_k$&$\Delta_k$ polynomial in&$\Delta_k^{(0)},\G_{n-1}^{(1)},\G_{n+1}^{(0)}$ polynomial in\\
    \hline
    (1)&$\Delta_{n-N-1}$&$z_{n-2N-1},\dots,z_{n-1}$&$\underline{c_{n-2N-1}},\dots,{c_{n-1}}$\\
    (2)&$\Delta_{n-N-2}$&$z_{n-2N-2},\dots,z_{n-2}$&$\underline{c_{n-2N-2}},\dots,{c_{n-2}}$\\
    (3)&$\vdots$&$\vdots$&$\vdots$\\ \hline
    (4)&$\Delta_{n-N}$&$z_{n-2N},\dots,z_{n}$&${c_{n-2N}},\dots,c_{n-1},\underline{a_-},\underline{a_{n+1}}$\\
    (5)&$\Delta_{n-N+1}$&$z_{n-2N+1},\dots,z_{n+1}$&${c_{n-2N+1}},\dots,c_{n+1},a_-,\underline{a_+},
              \underline{a}$\\
    (6)&$\Delta_{n-N+2}$&$z_{n-2N+2},\dots,z_{n+2}$&${c_{n-2N+2}},\dots,c_{n+2},a_\pm,a,\underline{c_{n+2}}$\\
    (7)&$\vdots$&$\vdots$&$\vdots$\\
    (8)&$\Delta_{n-1}$&$z_{n-N-1},\dots,z_{n+N-1}$&${c_{n-N-1}},\dots,c_{n-2},a_\pm,a$\\
       &&&$c_{n+2},\dots,\underline{c_{n+N-1}}$\\
    (9a)&$\G_{n-1}$& $x_{n-N-1},z_{n-N},\dots$ &$a_{n-N-2},c_{n-N-1},\dots$\\
        && $\dots,z_{n+N-2},x_{n+N-1}$ &$\dots,c_{n+N-1},\underline{a_{n+N}}$\\
    (9b)&$\G_{n+1}$&  $x_{n-N+1},z_{n-N+2},\dots$  &$a_{n-N+1},c_{n-N},\dots$\\
        &&$\dots,z_{n+N},x_{n+N+1}$&$\dots,c_{n+N-1},\underline{b_{n+N}},\crossed{a_{n+N+1}}$\\
    (10)&$\Delta_{n}$&$z_{n-N},\dots,z_{n+N}$&${c_{n-N-1}},\dots,a_{n-2},a_\pm$\\
       &&&$a_{n+2},\dots,\underline{c_{n+N+1}}$\\
    (11)&$\Delta_{n+2}$&$z_{n-N+2},\dots,z_{n+N+2}$&${c_{n-N+2}},\dots,a_{n-2},a_\pm$\\
        &&&    $a_{n+2},\dots,\underline{c_{n+N+2}}$\\
    (12)&$\vdots$&$\vdots$&$\vdots$\\
    \hline
  \end{tabular}
  \end{center}
\end{table}

\begin{proof}
We give the proof in the case $N>1$ only, leaving the case $N=1$ to the reader (see Remark \ref{N=1rem} for the
self-dual $N=1$ case). As in the self-dual case, we summarize the order in which we treat the different equations
in a table (see Table \ref{general_table}). The second column shows which $\Delta_k=(\G_k,\Gt_k)$ we consider. For
$k\neq n\pm1$ it is clear that each $\Delta_k$ appears (precisely once). The fact that $\G_{n-1}$ appears on line
$(9a)$, while $\Delta_{n-1}$ already appears on line $(8)$ comes from the fact that we consider in line $(9a)$ the
coefficient in $t$ of $\G_{n-1}(t)$ (rather than the coefficient in $t^0$); similarly, $\Gt_{n+1}$ is absent
because the nullity of $\Gt_{n+1}(0)$ is a consequence of the nullity of the other $\Delta_k(0)$ (Proposition
\ref{tang_gen_prop}). We know from Proposition~\ref{app_prop} that for any $k\in\Z$,
\begin{eqn}{sixpeight}
  &&\G_k(x,y;u)\in\R[x_{k-N},\dots,x_{k+N},y_{k-N+1},\dots,y_{k+N-1}],\\
  &&\Gt_k(x,y;u)\in\R[x_{k-N+1},\dots,x_{k+N-1},y_{k-N},\dots,y_{k+N}],
\end{eqn}
so that 
\begin{equation*}
  \Delta_k(x,y;u)\in\R[z_{k-N},\dots,z_{k+N}].  
\end{equation*}%
This leads, with no effort, to the third column of the table. For future use, let us recall that $\G_k$ depends
(linearly) on $x_{k-N}$ and on $x_{k+N}$, while $\Gt_k$ depends (linearly) on $y_{k-N}$ and on $y_{k+N}$.

\smallskip

Let us now turn, line by line, to the last column, which demands a careful inspection of the polynomials $\G_k$ and
$\Gt_k$. In particular, we show that these polynomials depend on the underlined parameter(s) (linearly), in such a
way that one can solve for them. In steps (1) -- (3) we have that $z_n$ is absent, so that $\Delta_{n-N-k}(0)$
($k\geq 1$) depends on $z_{n-2N-k}(0),\dots,z_{n-k}(0)$ only, i.e., on $c_{n-2N-k},\dots,c_{n-k}$. Now $\G_{n-N-k}$
depends on $x_{n-2N-k}$ (linearly), but not on $y_{n-2N-k}$, while the opposite is true for $\Gt_{n-N-k}$, so that
we can solve the equation $\G_{n-N-k}(0)=0$ linearly for $a_{n-2N-k}$, and similarly $\Gt_{n-N-k}(0)=0$ can be
solved linearly for $b_{n-2N-k}$ in terms of $c_{n-2N-k+1},\dots,c_{n-k}$. For $k=1$ this gives $a_{n-2N-1}$
(resp.\ $b_{n-2N-1}$) in terms of the $4N-1$ parameters $c_{n-2N},\dots,c_{n-2},a_{n-1}$, so that by taking
$k=2,3,\dots,$ we get recursively $c_{n-2N-k}$ in terms of these parameters, for all $k\geq 1$.

\smallskip

We now get to step (4) which is different because $\Delta_{n-N}$ involves $x_n$ and $y_n$. As for $\G_n$, according
to Proposition \ref{app_prop}, $x_n$ appears only in the leading term of $\G_{n-N}$, which we can write as
\begin{equation*}
  u_Nx_n\prod_{i=0}^{N-1}v_{n-N+i}=u_Nw_-\prod_{i=n-N}^{n-2}v_{i},\qquad u_N\neq0,
\end{equation*}%
where $w_-:=x_nv_{n-1}\in\A_n'$, as $w_-(t)=a_-a_{n-1}+O(t)$. Therefore, using~(\ref{sixpeight}),
\begin{equation*}
  \G_{n-N}(0)=u_Na_-a_{n-1}\prod_{i=n-N}^{n-2}\check c_i+\FF(a_{n-2N},c_{n-2N+1}, \dots,c_{n-1}),
\end{equation*}%
which can be solved linearly for $a_-$ in terms of the previous parameters ($\check c_i=1-a_ib_i\neq0$ for $n-N\leq
i\leq n-2$). Using the automorphism $\s$ (see (\ref{s_ext})),
\begin{equation*} 
  \Gt_{n-N}(0)=u_{-N}\frac{-a_-}{a_{n+1}}\prod_{i=n-N}^{n-2}\check c_i+ \FF(b_{n-2N},c_{n-2N+1},\dots,c_{n-1}),
\end{equation*}%
so that $ \Gt_{n-N}(0)=0$ can be solved linearly for $b_{n+1}=1/a_{n+1}$.

\smallskip

For step (5), $x_n$ and $y_n$ may be present in several terms in $\Delta_{n-N+1}$, but in view of Proposition
\ref{gen_pol_char_lemma}, $\G_{n-N+1}$ and $\Gt_{n-N+1}$ are polynomials in $z_{n-2N+1},\dots,$ $z_{n-1},z_{n+1}$
and in $w_1$ and $w_2$ and their $\sigma$ analogs only. Thus, $\G_{n-N+1}(0)$ and $\Gt_{n-N+1}(0)$ depend on their
leading terms only, to wit $c_{n-2N+1},\dots,c_{n-1},a_{n+1}$ and $a,a_\pm$. It follows that the only \emph{new}
parameters that appear at step (5) are $a_+$ and $a$. Let us show that they appear in such a way that we can solve
for them (linearly) in terms of the other parameters. We do this as in the self-dual case by isolating the leading
term in $\G_{n-N+1}$ as given in Proposition \ref{app_prop}, namely we write $\G_{n-N+1}$ as
\begin{equation}\label{step5}
  \G_{n-N+1}=-u_N(x_nw_1-x_{n+1})v_{n-1}\prod_{i=n-N+1}^{n-2}v_i+
      \FF(z_{n-2N+2},\dots,z_n),
\end{equation}
The relation (\ref{step5}) was obtained by writing the leading term
\begin{equation*}
  x_{n+1}v_n=x_{n+1}(1-x_ny_n)=x_{n+1}-(x_nw_1-x_n^2y_{n-1}),
\end{equation*}%
and throwing the $x_n^2y_{n-1}$ term into $\FF$.  Since $\G_{n-N+1}(t)=O(1)$ and since the first two terms in
(\ref{step5}) belong to $\A_n'$, the last term in (\ref{step5}) is also $O(1)$ in $t$; since in addition this term
does not contain $z_{n+1}$, by Proposition \ref{gen_pol_char_lemma} and (\ref{w_gen_def}) $x_n$ and $y_n$ can only
appear in it multiplied by $v_{n-1}=1-x_{n-1}y_{n-1}$, and so by Proposition \ref{general_laurent_prop} we may
conclude that the contribution from this term in $\G_{n-N+1}(0)$ will not involve $a_+$ or $a$. Also, the second
term in (\ref{step5}), $u_Nx_{n+1}v_{n-1}\prod_{i=n-N+1}^{n-2}v_i$ does not contribute to $\G_{n-N+1}(0)$ since
$v_{n-1}(t)=O(t)$ while all other factors are $O(1)$. Thus, the dependence on $a_+$ and $a$ in $\G_{n-N+1}(0)$
comes entirely from the first term in (\ref{step5}), which in view of Proposition \ref{general_laurent_prop} and
(\ref{w_gen_lau}) is given by
\begin{equation*} 
  \G_{n-N+1}(0)=-u_Na_-\Omega\prod_{i=n-N+1}^{n-2}\check c_i+\hbox{previous parameters}.
\end{equation*}%
By duality,
\begin{equation*}
   \Gt_{n-N+1}(0)=u_{-N}\frac{a_-a_{n-1}}{a_{n+1}}\sigma(\Omega) \prod_{i=n-N+1}^{n-2}\check c_i+\hbox{previous
            parameters},
\end{equation*}%
where $\sigma(\Omega)$ was given in (\ref{somega}). Since $\Omega$ and $\sigma(\Omega)$ are linearly independent,
as linear functions of $a_+$ and $a$, we can indeed solve $\Gt_{n-N+1}(0)=0$ and $\Gt_{n-N+1}(0)=0$ linearly for
$a_+$ and $a$ in terms of the other parameters.

\smallskip

Steps\footnote{Skip these steps if $N=2$.} (6) -- (8) are easy, the point being that by Proposition \ref{app_prop},
for $2\leq k\leq N-1$
\begin{equation*}
  \Delta_{n-N+k}(0)=
  \begin{matrix}{cc}
    u_N\\&u_{-N}  
  \end{matrix}
  c_{n+k}(v_{n-1}v_nv_{n+1})^{(0)}
  \renewcommand{\arraystretch}{0.5}
  \prod_{\begin{array}{c}
  \scriptstyle i=n-N+k\\
  \scriptstyle i\neq n-1,n,n+1
  \end{array}}^{n+k-1}\check c_i  \hbox{ + known.}
\end{equation*}%
Let us concentrate on the next steps, which are more exciting. In step (9a) we need to compute the linear term in
$\G_{n-1}(t)$, where we recall from Propositions \ref{tang_prop} and \ref{gen_pol_char_lemma} that
$\G_{n-1}\in\A_n'$, hence that this linear term only depends on the constant and linear terms of the elements of
$\G_{n-1}\in\A_n'$. Since $\G_{n-1}\in\R[x_{n-N-1},\dots,x_{n+N-1},y_{n-N},\dots,y_{n+N-2}]$, with leading term
\begin{equation*}
  \G_{n-1}=u_Nx_{n+N-1}\prod_{i=0}^{N-1}v_{n+i-1}+\cdots,
\end{equation*}%
we have from Proposition \ref{general_laurent_prop} that
\begin{equation*}
  \G_{n-1}(t)=\G_{n-1}^{(0)}+\left(u_Na_{n+N}(v_{n-1}v_nv_{n+1})^{(0)}\prod_{i=3}^{N}\check c_{n+i-1}+\cdots\right)t+O(t^2),  
\end{equation*}%
where the dots only involve previous parameters. Therefore we may solve $\G_{n-1}^{(1)}=0$ (linearly) for
$a_{n+N}$. Step (9b) is similar to step (9) in the self-dual case; notice that we postpone again $\Delta_n$ to the
next step. First of all $\G_{n+1}(t)=O(1)$ and so $\G_{n+1}\in \A_n'$. The leading term in $\G_{n+1}$, namely the
term $u_Nx_{n+N+1}v_{n+1}\prod_{i=1}^{N-1}v_{n+1+i}$ cannot contribute to $\G_{n+1}(0)$ because it is $O(t)$, which
explains the absence of $a_{n+N+1}$ in $\G_{n+1}(0)$. By Proposition \ref{app_prop}, $b_{n+N}$ can come only from
$y_{n+N}$, which appears only once, namely in
\begin{equation*}
  -u_{-N}y_{n+N}x_nx_{n+1}\prod_{i=0}^{N-2}v_{n+1+i}= -u_{-N}y_{n+N}x_{n+1}(x_nv_{n+1})\prod_{i=n+2}^{n+N-1}v_{i},
\end{equation*}%
yielding at $t=0$ a non-zero linear term in $b_{n+N}$, as $x_n(t)v_{n+1}(t)=O(1)$.

\smallskip

Step (10) is the hardest one, but we dealt with it in Lemma \ref{step10lemma}. Notice that after this step we have
that $\Delta_n(t)=O(t)$ since the nullity of the previous $\Delta_k(0)$ already implies that $\Delta_n(t)=O(1)$
(Proposition \ref{tang_prop}). Starting from step (11) everything goes smoothly, as $\Delta_k(t)=O(1)$ for $k>n+1$
and the leading term of $\G_k(0)$, resp.\ $\Gt_k(0)$ will produce precisely the new parameter $a_{k+N}$, resp.\
$b_{k+N}$ (linearly).
\qed
\end{proof}
%

%
\section{Singularity confinement}
We have constructed in the previous sections formal Laurent series for the Toeplitz lattice (in the self-dual and
general case) solving the recursion relations $\G_k(x(t);u(t))=0$ ($\Delta_k(x(t),y(t);u(t))=0$ in the general
case). We will now transform these into solutions of the recursion relations $\G_k(x;u)=0$ (resp.\
$\Delta_k(x,y;u)=0$), depending on a certain number of free parameters, and blowing up for only one (resp.\ two)
variables. We will mainly concentrate on the self-dual case, as the general case is dealt with in precisely the
same way.

The main tool to do this transformation is a formal version of the implicit function theorem, which we explain in
the case of one variable, the scalar case. Suppose that we have a formal series in $t$,
\begin{equation}\label{xstart}
  x(t;a)=a+f_1(a)t+f_2(a)t^2+\cdots;
\end{equation}
one may think for example of $x(t;a)$ as a formal solution of a vector field (differential equation $\dot x=F(x)$)
on the real line, with initial condition $x(0;a)=a$. In our case the functions $f_i$ will be rational. We wish
solve the equation $x(t;a)=\a$ formally, namely we wish to construct the formal series in $t$
\begin{equation*}
  a(t;\a)=\a+g_1(\a)t+g_2(\a)t^2+\cdots
\end{equation*}%
with the property that $x(t;a(t;\a))=\a$, as a formal $t$-series identity. Precisely, we claim that there exist for
any $s\in\N$ unique (rational) functions $g_1(\a),\dots,g_s(\a)$, such that
\begin{equation*}
  x(t;\a+g_1(\a)t+g_2(\a)t^2+\cdots+g_s(\a)t^s)-\a=O(t^{s+1}),
\end{equation*}%
where $x(t;\cdot)$ is given by (\ref{xstart}). This is a trivial consequence of a formal version of Taylor's
Theorem. For example, for $s=1$ we neglect all terms in $t^2$ and the condition on $g_1$ becomes
\begin{eqn*}
  x(t;\a+g_1(\a)t)-\a+O(t^2)&=&g_1(\a)t+f_1(\a+g_1(\a)t)t+O(t^2)\\
    &=&(g_1(\a)+f_1(\a))t+O(t^2),
\end{eqn*}%
so that $g_1(\a)=-f_1(\a)$. For $s=2$ we neglect the terms in $t^3$, giving
\begin{eqn*}
  \lefteqn{x(t;\a-f_1(\a)t+g_2(\a)t^2)-\a+O(t^3)}\hskip0.11cm\\
    &=&-f_1(\a)t+g_2(\a)t^2+f_1(\a-f_1(\a)t)t+f_2(\a)t^2+O(t^3)\\
    &=&g_2(\a)t^2+f_1'(\a)(-f_1(\a)t)t+f_2(\a)t^2+O(t^3)\\
    &=&(g_2(\a)-f_1(\a)f'_1(\a)+f_2(\a))t^2+O(t^3),
\end{eqn*}%
which has $g_2(\a):=f_1(\a)f'_1(\a)-f_2(\a)$ as a unique solution. Continuing in this way it is clear that $g_i(\a)$
equals $-f_i(\a)$, up to a differential polynomial in the $f_j(\a)$, with $j<i$. Notice that when all $f_i(a)$ are
rational function the same will be true for all $g_j(\a)$.

\smallskip

Let us apply this to the formal Laurent series that we have constructed for the self-dual Toeplitz lattice, and
that yield formal solutions to the recursion relations $\G_k(t):=\G_k(x(t);u(t))=0$, where $k\in\Z$. Recall from
Proposition \ref{sd_restr_prop} that these formal Laurent solutions $x_k(t)$ depend on $2N-1$ parameters
$a_{n-2N},\dots,a_{n-2}$, which are the leading coefficients of $x_{n-2N},\dots,x_{n-2}$, namely
\begin{equation}\label{toinvert}
  x_k(t)=a_k+O(t),\qquad k=n-2N,\dots,n-2,
\end{equation}
where the higher order terms are rational functions of the parameters $a_{n-2N},\dots,$ $a_{n-2}$. Besides the
parameters $a_k$ these functions also depend (polynomially) on the parameters $u=(u_1,\dots,u_N)$ that define the
recursion relations, namely $x_k(t)=x_k(t;a_{n-2N},\dots,a_{n-2};u)$, for $n-2N\leq k\leq n-2$. The formal implicit
function theorem then leads to the following proposition.
\begin{proposition}\label{par_sub_prop}
  There exist for $k=n-2N,\dots,k=n-2$ rational functions
  \begin{equation*}
    a_k^{(i)}=a_k^{(i)}(\a_{n-2N},\dots,\a_{n-2};u_1,\dots,u_N)
  \end{equation*}%
  such that $a_k:=\sum_{i=0}^\infty a_k^{(i)}t^i$, $k=n-2N,\dots,n-2$ formally inverts (\ref{toinvert}), i.e.,
  \begin{equation*}
     x_k\left(t;\sum_{i=0}^\infty a_{n-2N}^{(i)}t^i,\dots, \sum_{i=0}^\infty a_{n-2}^{(i)}t^i;u\right)=\a_k,
  \end{equation*}%
  for $k=n-2N,\dots,n-2$, with $a_k^{(0)}=\a_k$.
\end{proposition}
\qed

We can use these series to replace the free parameters $a_{n-2N},\dots,a_{n-2}$ in the series $x_k(t),\,k\in\Z$, by
$\a:=(\a_{n-2N},\dots,\a_{n-2})$, where we think of the latter as (partial) initial conditions to the recursion
relation. To do this, one simply substitutes $a_k=\sum_{i=0}^\infty a_k^{(i)}t^i$ for $k=n-2N,\dots,n-2$ in each of
the series $x_k(t)=x_k(t;a_{n-2N},\dots,a_{n-2};u)$, and rewrites this as a series in $t$; by construction, this
simply gives $x_k(t)=\a_k$ for $k=n-2N,\dots,k=n-2$.  For $k=n-1$, this yields
\begin{equation*}
  x_{n-1}(t)=\e+\sum_{i=1}^\infty x_{n-1}^{(i)}(a;u)t^i =\e+\sum_{i=1}^\infty\xi_{n-1}^{(i)}(\a;u)t^i,
\end{equation*}%
where we recall that $\e^2=1$. The functions $\xi_{n-1}^{(i)}$ are rational in $\a$ and $u$. We will now use the
formal implicit\footnote{Call this the formal inverse function theorem, if you wish.} function theorem again, but
in a form which is different from the one explained above: putting $x_{n-1}(t)=\e+\l(t)$, i.e., we put
\begin{equation*}
  \l:=\sum_{i=1}^\infty \xi^{(i)}(\a;u)t^i,
\end{equation*}%
which we solve for $t$ as a formal series in $\l$, 
\begin{equation}\label{tgo}
  t(\l)=\sum_{i=1}^\infty \tau^{(i)}(\a;u)\l^i,
\end{equation}
where it is important to note that the constant term in this series is absent. Indeed, let us first substitute
(\ref{tgo}) in the series for $a_k$ that was obtained in Proposition (\ref{par_sub_prop}), to get
$a_k=a_k(\a;\l;u)$. Then, the latter and $t(\l)$ are substituted in all $x_k(t)$, to yield series in $\l$ whose
coefficients are rational functions of $\a=(\a_{n-2N},\dots,\a_{n-2})$ (and of $u=(u_1,\dots,u_N)$), which take the
following form.
\begin{equation*}
  \renewcommand{\arraystretch}{1.7}
  \begin{array}{rclrcl}
  x_k(\l,\a;u)&=&\sum_{i=0}^\infty \chi_k^{(i)}(\a;u)\l^i,\qquad &k&<&n-2N,\\
  x_k(\l,\a;u)&=&\a_k,&n-2N&\leq& k<n-1,\\
  x_{n-1}(\l,\a;u)&=&\e+\l,\\
  x_n(\l,\a;u)&=&\frac1\l\sum_{i=0}^\infty \chi_n^{(i)}(\a;u)\l^i,\\
  x_{n+1}(\l,\a;u)&=&-\e+\sum_{i=1}^\infty \chi_{n+1}^{(i)}(\a;u)\l^i,\qquad\\
  x_k(\l,\a;u)&=&\sum_{i=0}^\infty \chi_k^{(i)}(\a;u)\l^i,\qquad &n+1&<&k.
  \end{array}
\end{equation*}%
It may seem that we have reached the final result, but we should not forget that these series are constructed from
solutions $x=x(t)$ to the recursion relations $\G_k(x;u(t))$, where $u(t)=(u_1+t,u_2,\dots,u_N)$. However, letting
$U=(U_1,\dots,U_n):=u(t)$, and using (\ref{tgo}) to get rid of $t$, we have that
\begin{equation*}
  x_k(\l,\a;(U_1-t(\l),U_2,\dots,U_N)),\ k\in\Z\ \hbox{ solves }\ \G_k(x;U),\ k\in\Z.
\end{equation*}%
Notice that, when it is all worked out, the $x_k$ are formal power series in $\l$ (except $x_n$ which has a simple
pole in $\l$), and their coefficients are rational functions of the initial conditions $\a_{n-2N},\dots,\a_{n-2}$
and of the parameters $U_1,\dots,U_n$. Writing
\begin{eqnarray*}
  x_k(\l,\a;(U_1-t(\l),U_2,\dots,U_N))&=&\sum_{i=0}^\infty x_k^{(i)}(\a;U)\l^i,\
  k\in\Z\setminus\set{n}\\
  x_n(\l,\a;(U_1-t(\l),U_2,\dots,U_N))&=&\sum_{i=-1}^\infty x_n^{(i)}(\a;U)\l^i,\
\end{eqnarray*}
leads to our final result.

\begin{theorem}
  The recursion relations $\G_k(x;U)=0,\,k\in\Z$ admit for any $n\in\Z$ two\footnote{parametrized by
  $\epsilon=\pm1.$} formal Laurent solution $x=(x_k(\a,\l;U))_{k\in\Z}$, depending on $2N$ free parameters
  $\a=(\a_{n-2N},\dots,\a_{n-2})$ and $\l$ with $x_n$ having a (simple) pole for $\l\to 0$, and no other
  singularities. Explicitly, these series with coefficients rational in $\a$  are given by
\begin{equation*}
  \renewcommand{\arraystretch}{1.7}
  \begin{array}{rclrcl}
  x_k(\l,\a;U)&=&\sum_{i=0}^\infty x_k^{(i)}(\a;U)\l^i,\qquad &k&<&n-2N,\\
  x_k(\l,\a;U)&=&\a_k,&n-2N&\leq& k<n-1,\\
  x_{n-1}(\l,\a;U)&=&\e+\l,\\
  x_n(\l,\a;U)&=&\frac1\l\sum_{i=0}^\infty x_n^{(i)}(\a;U)\l^i,\\
  x_{n+1}(\l,\a;U)&=&-\e+\sum_{i=1}^\infty x_{n+1}^{(i)}(\a;U)\l^i,\qquad\\
  x_k(\l,\a;U)&=&\sum_{i=0}^\infty x_k^{(i)}(\a;U)\l^i,\qquad &n+1&<&k.
  \end{array}
\end{equation*}%
\end{theorem}
The corresponding theorem for the recursion relations $\Delta_k=0$, which was formulated in the introduction
(Theorem \ref{intro_thm}) follows in the same way, using the formal Laurent solutions $z(t)$ that solve the recursion
relations.

%
\section{Appendix}

In this appendix we obtain the leading terms of the polynomials $\G_k$ and~$\Gt_k$, which are needed in Sections
\ref{conf_self-dual_sec} and \ref{conf_general_sec}. The notations are as in the body of the paper, namely $P_1$
and $P_2$ are polynomials of degree $N$ (see (\ref{P_def})), the matrices $L_1$ and $L_2$ are defined by
(\ref{lax_mats}) and the polynomials $\G_k$ and $\Gt_k$ are defined by (\ref{gam_def}). Since $\G_k$ is given by
\begin{equation}\label{Gk_app}
    \G_k(x,y;u):=\displaystyle\frac{v_k}{y_k}\left(
    \renewcommand{\arraystretch}{1.3}
    \begin{array}{c}
      -(L_1P'_1(L_1))_{k+1,k+1}-(L_2P_2'(L_2))_{k,k}\\
      +(P'_1(L_1))_{k+1,k}+(P_2'(L_2))_{k,k+1}
    \end{array}
    \right)+kx_k,
\end{equation}%
we need, by duality, only to determine the leading terms of $(L_1^s)_{kk}$ and of $(L_1^s)_{k+1,k}$, for
$s,k\in\Z$, with $s\geq2$, which will be done in the following lemma. Notice that the leading terms of $\Gt_k$ will
also follow from it, by duality.
\begin{lemma}\label{app_lemma}
  For $k\in\Z$ and $s\in\N$, with $s\geq2$, the diagonal and first subdiagonal entries of the Toeplitz matrices
  $L_1$ and $L_2$, defined in (\ref{lax_mats}), are polynomials in the following variables,
  \begin{eqnarray*}
    \left(L_1^{s}\right)_{kk}&\in&\R[x_{k-s+1},\dots,x_{k+s-1},y_{k-s}, \dots,y_{k+s-2}],\\
    \left(L_1^{s}\right)_{k+1,k}&\in&\R[x_{k-s+1},\dots,x_{k+s},y_{k-s}, \dots,y_{k+s-1}].
  \end{eqnarray*}
  More precisely\footnote{We give in each case the terms that will be used, no more, no less. When $s=2$ only the
  first two lines survive; the term on fourth line coincides with the first term on the second line and should only
  be counted once.},
  \begin{eqnarray*}
    \left(L_1^{s}\right)_{kk}
         &=&-\,x_{k+s-1}y_{k-1}\prod_{i=1}^{s-1}v_{k+i-1}+x_{k+s-2}^2y_{k+s-3}y_{k-1}\prod_{i=1}^{s-2}v_{k+i-1}\\
         &&-\,x_{k+s-2}\left(y_{k-2}v_{k-1}-2y_{k-1}\sum_{j=1}^{s-2}x_{k+j-1}y_{k+j-2}\right)
                   \prod_{i=1}^{s-2}v_{k+i-1}\\
      &&+\,\FF_1(x_{k-s+2},\dots,x_{k+s-3},y_{k-s+1},\dots,y_{k+s-3})\\
      &&-\,x_ky_{k-s}\prod_{i=1}^{s-1}v_{k-i}
  \end{eqnarray*}
and
\begin{eqnarray*}
    \left(L_1^{s}\right)_{k+1,k}&=&-\,x_{k+s}y_{k-1}\prod_{i=1}^{s-1}v_{k+i}-x_{k+1}y_{k-s}\prod_{i=1}^{s-1}v_{k-i}\\
     &&+\,\FF_2(x_{k-s+2},\dots,x_{k+s-1},y_{k-s+1},\dots,y_{k+s-2})
\end{eqnarray*}
  where $\FF_1$ and $\FF_2$ are polynomials in their arguments.
\end{lemma}
\begin{proof}
The following notation is useful for obtaining formulas of this type. To the bi-infinite vector $x$ we associate,
for any $k\in\Z$ a bi-infinite diagonal matrix $X^{(k)}$ by putting $X_{ij}^{(k)}=x_{i+k}\delta_{ij}$ (Kronecker
delta). Similarly we introduce the diagonal matrices $Y^{(k)}$ and $V^{(k)}$, associated to $y$ and $v$. We denote
by $\Delta$ the shift operator, which we view as a bi-infinite matrix, with entries
$\Delta_{ij}:=\delta_{i+1,j}$. It is easy to verify that
\begin{equation*}
  \Delta^iX^{(j)}=X^{(i+j)}\Delta^i,\qquad i,j\in\Z,
\end{equation*}%
which is the main formula that we will use, as it allows us to push all $\Delta$ to the right (or to the left). One
obvious consequence is that a monomial in $X,Y,V$ and $\Delta$ will only have a non-zero diagonal when it is
independent of $\Delta$ (i.e., the sum of all powers of $\Delta$ is zero). In order to apply this to obtain the
above formulas, observe that $L_1$ and $L_2$ can be written as
\begin{eqnarray*}
  L_1&=&\Delta V^{(-1)}-\sum_{i\geq0}\Delta^{-i}X^{(i)}Y^{(-1)}=V^{(0)}\Delta-\sum_{i\geq0}X^{(0)}Y^{(-i-1)}\Delta^{-i},\\
  L_2&=&\Delta^{-1}V^{(0)}-\sum_{i\geq0}\Delta^{i}X^{(-i-1)}Y^{(0)}=V^{(-1)}\Delta^{-1}-\sum_{i\geq0}X^{(-1)}Y^{(i)}\Delta^{i}.
\end{eqnarray*}
Notice that, in view of what we said, all diagonal entries of $(V^{(0)}\Delta)^{s-1}$ are zero. Therefore, it follows
from the second formula for $L_1$ that the leading term in $x$ of the diagonal terms of $L_1^{s}$ will be gotten
from the product
\begin{equation}\label{app_prod}
  -(V^{(0)}\Delta)^{s-1}\sum_{i\geq0}X^{(0)}Y^{(-i-1)}\Delta^{-i}.
\end{equation}
The diagonal entries of (\ref{app_prod}) are obtained by taking $i=s-1$,
which yields
\begin{eqnarray*}
  \left(-(V^{(0)}\Delta)^{s-1}X^{(0)}Y^{(-s)}\Delta^{-s+1}\right)_{kk}
   &=&-\left(V^{(0)}\dots V^{(s-2)} X^{(s-1)}Y^{(-1)}\right)_{kk}\\
  &=&-x_{k+s-1}y_{k-1}\prod_{i=1}^{s-1}v_{k+i-1}.
\end{eqnarray*}
Notice that this leading term already contains $x_{k+s-2}$, and that it yields, through
$v_{k+s-2}=1-x_{k+s-2}y_{k+s-2}$, the single term that contains $y_{k+s-2}$, which is the highest $y$ variable that
appears in $(L_1^s)_{kk}$.

\smallskip

In order to get the other terms in $L_1^s$ that lead to $x_{k+s-2}$ we need $\Delta^{s-2}$ in front of $X^{(0)}$,
i.e., we need $s-2$ copies of $V^{(0)}\Delta$ (not necessarily consecutive), on the left of $-\sum_{i\geq0}X^{(0)}
Y^{(-i-1)}\Delta^{-i}$. For the remaining factor we can have another copy of $V^{(0)}\Delta$ or of
$-\sum_{i\geq0}X^{(0)}Y^{(-i-1)}\Delta^{-i}$, inserted at an arbitrary place inside the product
$-(V^{(0)}\Delta)^{s-2}\sum_{i\geq0}X^{(0)} Y^{(-i-1)}\Delta^{-i}$. This leads to three possible types of
terms. For the first one, we put another $V^{(0)}\Delta$ at the end
\begin{equation*}
  -(V^{(0)}\Delta)^{s-2}\sum_{i\geq0}X^{(0)}Y^{(-i-1)}\Delta^{-i}(V^{(0)}\Delta),
\end{equation*}%
and we get the $k,k$ diagonal term  by taking $i=s-1$, which gives
\begin{equation*}
  \left(-(V^{(0)}\Delta)^{s-2}X^{(0)}Y^{(-s)}\Delta^{1-s}V^{(0)}\Delta\right)_{kk}=-x_{k+s-2}y_{k-2}\prod_{i=0}^{s-2}v_{k+i-1}.
\end{equation*}%
For the second one we put another $-\sum_{j\geq0}X^{(0)}Y^{(-j-1)}\Delta^{-j}$ at the end,
\begin{equation*}
  (V^{(0)}\Delta)^{s-2}\sum_{i\geq0}X^{(0)}Y^{(-i-1)}\Delta^{-i} \sum_{j\geq0}X^{(0)}Y^{(-j-1)}\Delta^{-j};
\end{equation*}%
its diagonal terms are given by taking $i+j=s-2$, i.e., from
\begin{equation*}
  (V^{(0)}\Delta)^{s-2}\sum_{j=0}^{s-2}X^{(0)}Y^{(j-s+1)} X^{(j-s+2)}Y^{(-s+1)}\Delta^{2-s},  
\end{equation*}%
whose $k,k$ term is given by
\begin{equation*}
  y_{k-1}\left(x_{k+s-2}^2y_{k+s-3}+x_{k+s-2}\sum_{j=0}^{s-3}x_{k+j}y_{k+j-1}\right) \prod_{i=1}^{s-2}v_{k+i-1}.
\end{equation*}%
The third term has been  obtained by inserting the constant term $-X^{(0)}Y^{(-1)}$ of
$-\sum_{j\geq0}X^{(0)}Y^{(-j-1)}\Delta^{-j}$ at all possible places in the product $(V^{(0)}\Delta)^{s-2}$, namely
from
\begin{equation*} 
  \sum_{j=0}^{s-3}(V^{(0)}\Delta)^j(X^{(0)}Y^{(-1)})(V^{(0)}\Delta)^{s-j-2}\sum_{i\geq0}X^{(0)}Y^{(-i-1)}\Delta^{-i},
\end{equation*}%
with $i=s-2$, so that its $k,k$ term is given by
\begin{equation*}
 \left(y_{k-1}x_{k+s-2} \sum_{j=0}^{s-3}x_{k+j}y_{k+j-1}\right)\prod_{i=1}^{s-2}v_{k+i-1},
\end{equation*}%
which, combined with the first two terms, yields the leading terms of $(L_1^s)_{kk}$. Using the first formula for
$L_1$, the lowest term in $y$ of the diagonal terms of $L_1^{s}$ is gotten from
\begin{equation*}
  -\Delta^{-s+1}X^{(s-1)}Y^{(-1)}(\Delta V^{(-1)})^{s-1}=-X^{(0)}Y^{(-s)}V^{(-s+1)}\dots V^{(-1)},
\end{equation*}%
whose $k,k$ entry is $-x_ky_{k-s}\prod_{i=1}^{s-1} v_{k-i}$. It contains the lowest term in $x$, through
$v_{k-s+1}=1-x_{k-s+1}y_{k-s+1}$.

\goodbreak

One obtains similarly the entries of $(L_1^s)_{k+1,k}$ by selecting the terms in $L_1^s$ that contain precisely
$\Delta^{-1}$. Notice in this respect that if $M$ is a bi-infinite diagonal matrix then
$(M\Delta^{-1})_{k+1,k}=M_{k+1,k+1}$. It follows that the leading term in $x$ of $(L_1^s)_{k+1,k}$, which contains
also the leading term in $y$, is obtained from the product (\ref{app_prod}), with $i=s$, yielding
\begin{equation*}
  -\left(V^{(0)}\dots V^{(s-2)} X^{(s-1)}Y^{(-2)}\right)_{k+1,k} =-x_{k+s}y_{k-1}\prod_{i=1}^{s-1}v_{k+i}.
\end{equation*}%
The lowest term in $y$, which contains the lowest term in $x$, is obtained in the same way.
\qed
\end{proof}
The above lemma and (\ref{Gk_app}) lead by direct substitution to the following proposition.
\begin{proposition}\label{app_prop}
  For $k\in\Z$, the polynomials $\G_k$ and $\Gt_k$ depend on the following variables $x_i$ and $y_i$:
  \begin{eqnarray*}
    &&\G_k(x,y;u)\in\R[x_{k-N},\dots,x_{k+N},y_{k-N+1},\dots,y_{k+N-1}],\\
    &&\Gt_k(x,y;u)\in\R[x_{k-N+1},\dots,x_{k+N-1},y_{k-N},\dots,y_{k+N}].
  \end{eqnarray*}
  More precisely\footnote{As in the case of Lemma \ref{app_lemma}, when $N=2$ then the term
  $-u_2x_kx_{k+1}
y_{k-1}v_k$, which appears twice, should only be taken into account once.}, 
  \begin{eqnarray*}
    \G_k(x,y;u)&=&u_Nx_{k+N}\prod_{i=0}^{N-1}v_{k+i}-u_Nx_{k+N-1}^2y_{k+N-2} \prod_{i=0}^{N-2}v_{k+i}\\
      &&\,-u_Nx_{k+N-1}\left(x_ky_{k-1}+2\sum_{j=1}^{N-2}x_{k+j}y_{k+j-1}\right)\prod_{i=0}^{N-2}v_{k+i}\\ 
      &&+\,(u_{N-1}x_{k+N-1}-u_{-N}y_{k+N-1}x_{k-1}x_k)\prod_{i=0}^{N-2}v_{k+i}\\
      &&+\,v_k\FF(x_{k-N+1},\dots,x_{k+N-2},y_{k-N+2},\dots,y_{k+N-2})+kx_k\\
      &&-\,(u_Nx_kx_{k+1}y_{k-N+1}-u_{-N}x_{k-N}v_{k-N+1})\prod_{i=0}^{N-2}v_{k-i},
  \end{eqnarray*}
  where $\FF$ is a polynomial in its arguments, with a similar statement for $\Gt_k$ gotten by duality. In the
  self-dual case, $\G_k$ takes the simpler form
  \begin{eqnarray*}
    \G_k(x;u)&=&u_Nx_{k+N}\prod_{i=0}^{N-1}v_{k+i}+u_{N-1}x_{k+N-1}\prod_{i=0}^{N-2}v_{k+i}\\
       &&\,-u_Nx_{k+N-1}\left(x_{k+N-1}x_{k+N-2}+ 2\sum_{j=0}^{N-2}x_{k+j}x_{k+j-1}\right)\prod_{i=0}^{N-2}v_{k+i}\\
       &&+\,v_k\FF(x_{k-N+1},\dots,x_{k+N-2})+kx_k\\
       &&-\,u_N(x_kx_{k+1}x_{k-N+1}-x_{k-N}v_{k-N+1})\prod_{i=0}^{N-2}v_{k-i}.
  \end{eqnarray*}
\end{proposition}
\bibliographystyle{abbrv} 
\def\cprime{$'$}

%
\end{document}